\documentclass[review]{elsarticle}
\usepackage{graphics}
\usepackage{graphicx}
\usepackage{float}%place images here with [H]
\usepackage{bm}
\usepackage[nomarkers,notablist,nofiglist]{endfloat}
\usepackage{lineno,hyperref}
\modulolinenumbers[5]
\usepackage{siunitx}         
\graphicspath{ {./Figures/} } 
\usepackage{soul}
\usepackage{amsmath}
\usepackage{geometry} % to change the page dimensions
\usepackage{color} 
\usepackage{outlines}
\usepackage{textcomp}
\journal{arXiv}

\begin{document}

\begin{frontmatter}

	\title{Deconvoluting Thermomechanical Effects in X-ray Diffraction Data using Machine Learning}
	%% Group authors per af{}filiation:

	%% or include af{}filiations in footnotes:
	\author[1,3]{Rachel E. Lim}
	\author[1]{Shun-Li Shang}
	\author[2]{Chihpin Chuang}
	\author[3]{Thien Q. Phan}
	\author[1]{Zi-Kui Liu}
	\author[1]{Darren C. Pagan\corref{mycorrespondingauthor}}
	\ead{dcp5303@psu.edu}

	\cortext[mycorrespondingauthor]{Corresponding author}

	\address[1]{Pennsylvania State University, University Park, PA 16802, USA}
 	\address[2]{Argonne National Laboratory, Lemont, IL 60439, USA}
	\address[3]{Lawrence Livermore National Laboratory, Livermore, CA 14850, USA}

	\begin{abstract}
        X-ray diffraction is ideal for probing sub-surface state during complex or rapid thermomechanical loading of crystalline materials. However, challenges arise as the size of diffraction volumes increases due to spatial broadening and inability to deconvolute the effects of different lattice deformation mechanisms. Here, we present a novel approach to use combinations of physics-based modeling and machine learning to deconvolve thermal and mechanical elastic strains for diffraction data analysis. The method builds on a previous effort to extract thermal strain distribution information from diffraction data. The new approach is applied to extract the evolution of thermomechanical state during laser melting of an Inconel 625 wall specimen which produces significant residual stress upon cooling. A combination of heat transfer and fluid flow, elasto-plasticity, and X-ray diffraction simulations are used to generate training data for machine-learning (Gaussian Process Regression, GPR) models that map diffracted intensity distributions to underlying thermomechanical strain fields. First-principles density functional theory is used to determine accurate temperature-dependent thermal expansion and elastic stiffness used for elasto-plasticity modeling. The trained GPR models are found to be capable of deconvoluting the effects of thermal and mechanical strains, in addition to providing information about underlying strain distributions, even from complex diffraction patterns with irregularly shaped peaks.

	\end{abstract}


\end{frontmatter}

\section{Introduction}

Understanding and controlling the development of residual stress during traditional welding, and now additive manufacturing (AM), is an ongoing challenge. Rapid, localized heating then solidification leads to large thermal stresses which, in turn, lead to plastic flow and residual stress. These residual stresses can drive hot-cracking or cause such severe distortions that a part is unusable. Optical and thermography measurements provide a means to characterize the temperature gradients driving stress development \cite{moylan2014infrared,everton2016review,fox2017measurement,fisher2018toward,montazeri2019heterogeneous,dunbar2018assessment,forien2020detecting,ashby2022thermal}, but are limited to the surface. In addition, while there has been a significant focus placed on developing high-speed X-ray imaging at synchrotron sources to monitor melt pool characteristics and the development of porosity in support of AM processing, the use of scattering and the study of stress development has been limited in comparison. However, recent efforts are building upon imaging efforts to also use scattering to study complex rapid solidification processes \cite{Kenel2016,Calta2018,Cunningham2019,Hocine2020,oh2021high,oh2021microscale,thampy2020subsurface,silveira2023microstructure,chen2023quantitative,scheel2023close,dass2023dendritic}. More specifically, diffraction processes are capable of monitoring the development of texture, phases, and residual strains (and related stress).

An acute challenge in using diffraction to monitor stress development in combined thermal and mechanical loading is that both temperature and mechanical stress alter the state of the crystal lattice (which is what is probed during diffraction measurements). Specifically in cubic crystal types, both temperature and hydrostatic stress will alter the volumetric portion of the lattice strain tensor in the same fashion, making deconvolution of the effects difficult. Assumptions can be made regarding whether the thermal or mechanical alterations to lattice state are dominant for analysis, but these can introduce significant uncertainty into interpretation of data when magnitudes become approximately equal. In addition, the spread of thermal and mechanical strain \emph{distributions} through a diffraction volume can be extensive, particularly when using high-energy X-rays in transmission through thick specimens. A single diffraction image (projection) does not provide sufficient information for direct reconstruction of the distributions present.

To overcome these challenges, we present a machine learning (ML) approach in which Gaussian Process Regression (GPR) models are trained to learn the relationship between diffraction patterns underlying thermomechanical strain distributions. The ML training process is supported by physics-based modeling of the heating and cooling processes that leads to stress development. This modeling, along with accurate diffraction simulations, is used to create a `dictionary' of diffraction patterns for ML model training. The trained ML model is then \emph{transferred} to the analysis of experimental data. To attain accurate thermal and mechanical properties, density functional theory (DFT) is utilized. This work builds on our previous effort to extract temperature (thermal strain) distribution metrics from \emph{in situ} diffraction patterns \cite{Lim:xx5027}. However, as mentioned, due to convolution of the effects of thermal and mechanical strain, this previous effort was only accurate at high temperatures. Transfer learning is particularly appropriate for cases where the underlying physics is fairly well understood and modeled. Any uncertainty in the accuracy of the model may have implications to the ``transferring" to real data. Thus, high fidelity models are required and the improvements to our previous work have been made through the following advances: i) the use of DFT for temperature-dependent thermomechanical properties, ii) modeling of stress distributions using a thermomechanical elasto-plasticity model, iii) update of the X-ray diffraction modeling framework to incorporate anisotropic lattice strains during mechanical loading, and iv) update of the GPR model training to utilize anisotropic diffraction ring expansion and contraction.

\begin{figure}[h!]
    \centering \includegraphics[width=0.9\textwidth]{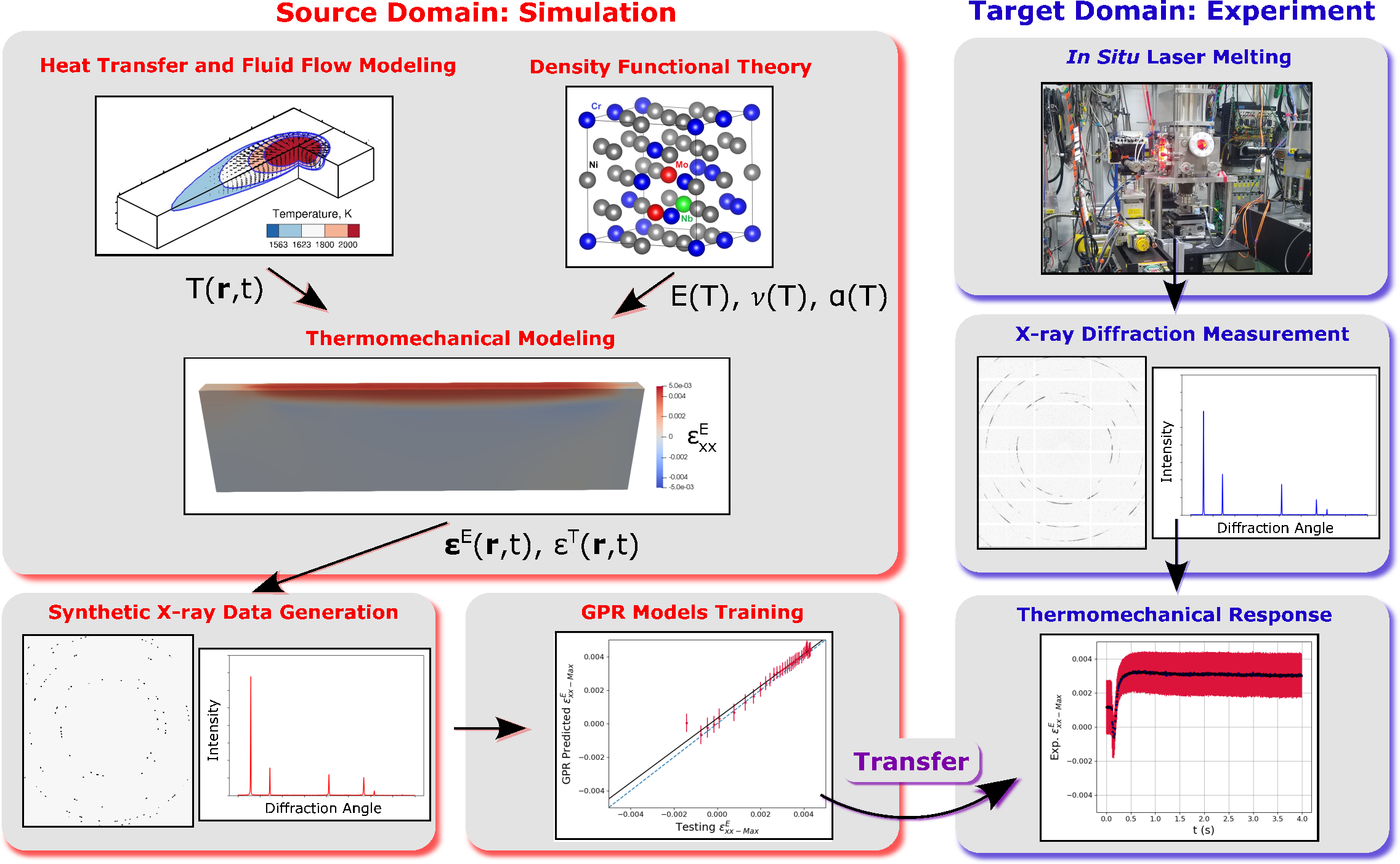}
    %\vspace{-0.15in}
    \caption{Overview of the simulations used to train Gaussian Process Regression (GPR) models for X-ray diffraction data analysis. The trained models are subsequently transferred to the target (experimental) domain to separate thermal and mechanical strain effects from experimental X-ray diffraction data. Temperature fields $T(\bm{r},t)$ and temperature-dependent properties (Young's modulus $E(T)$, Poisson's ratio $\nu(T)$, and coefficient of thermal expansion $\alpha(T)$)  are used as input for thermomechanical modeling to predict evolving thermal $\varepsilon^T(\bm{r},t)$ and elastic strains $\bm{\varepsilon}^E(\bm{r},t)$. These strains are used for input into X-ray diffraction simulations and GPR model training. The  trained GPR models are then used for analysis of experimental data.}
    \label{fig:overview}
\end{figure}

The approach is demonstrated through extraction of thermal and elastic strain distribution metrics from a wall specimen of Inconel 625 in a transmission geometry that has become a standard for \emph{in situ} AM experiments. An overview of the data-processing workflow is provided in Fig. \ref{fig:overview}. This work begins by describing the experimental data to be analyzed in \S \ref{sec:expdesc}. Next in \S \ref{sec:tdata}, summaries of the various physics models used for GPR model training are provided. The GPR models used and their training is described  in \S \ref{sec:training}. The trained GPR models are then applied to experimental \emph{in situ} diffraction data in \S \ref{sec:results}. The work finishes by discussing our approach and avenues for future study in \S \ref{sec:disc}.

\section{Material and Experiment Description}
\label{sec:expdesc}

The material tested was Inconel 625 (IN625) that was extracted from a block built using laser powder bed fusion at the National Institute of Standards and Technology (NIST) \cite{levine2020outcomes}. The block was built in an EOS M290 system using powder also obtained from EOS. The block dimensions were 50 mm $\times$ 15 mm $\times$ 6 mm where the 6 mm direction is the build direction. The build followed manufacturer recommendations \cite{son2020creep} of laser power of 285 W, laser speed of 960 mm/s, and interlayer rotation of 67.5$^\circ$. After build, the specimen was stress-relief heat treated at 800~$^\circ$C for 2 hr. A wall specimen was then extracted using electro-discharge machining with dimensions of 15 mm $\times$ 0.53 mm $\times$ 3 mm, with the 3 mm dimension being aligned with the build dimension. Previous characterization (including electron backscatter diffraction) of the material found that the grain size ranged from approximately 10 $\mu$m to 100 $\mu$m with a mean between 25 $\mu$m and 30 $\mu$m \cite{Lim:xx5027}. 

\begin{figure}[h]
      \centering \includegraphics[width=0.75\textwidth]{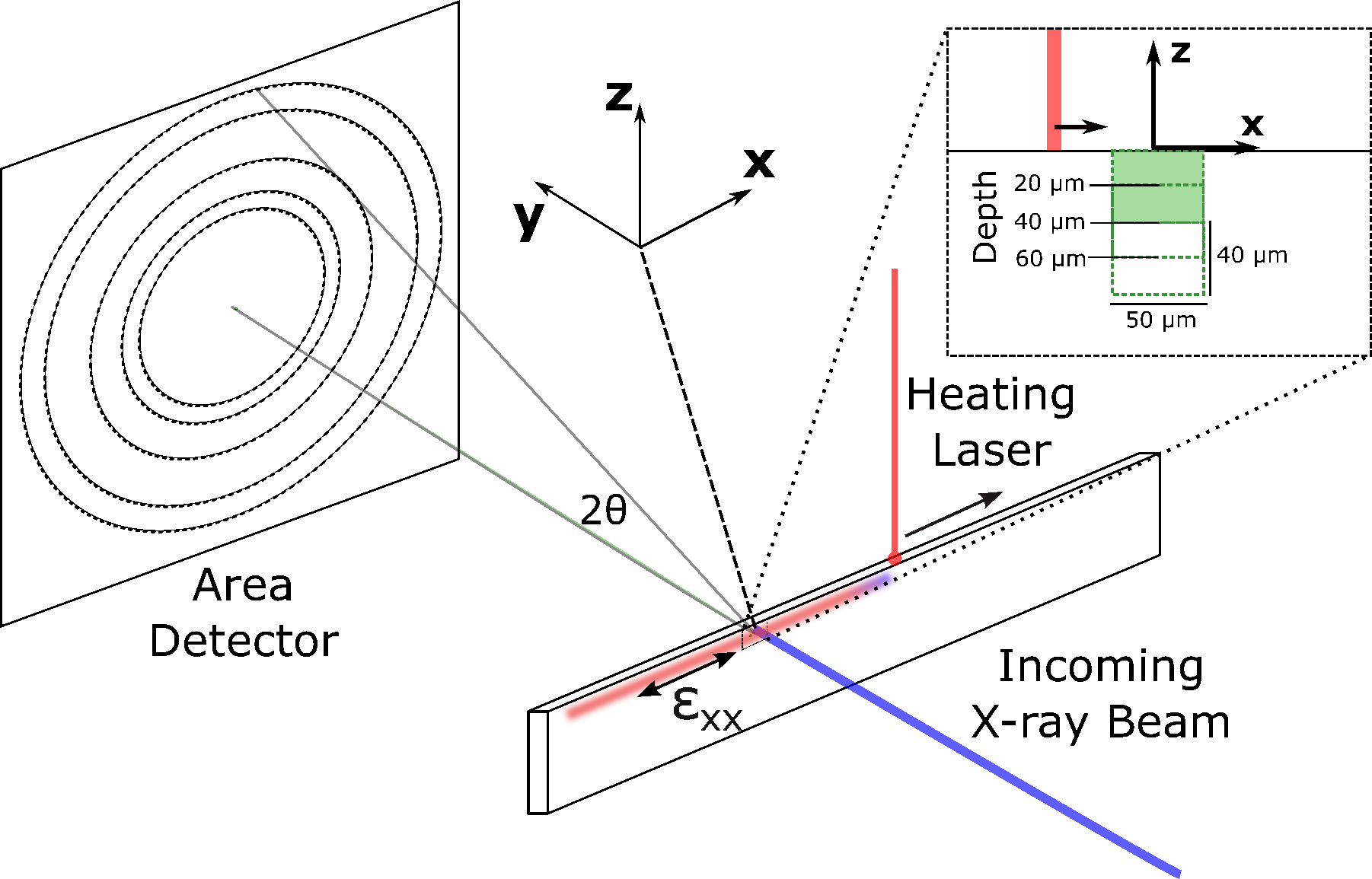}
      %\vspace{-0.15in}
      \caption{Schematic of the (simulation and experiment) geometry for the wall laser melting experiment. The inset indicates points where diffraction measurements and simulations were performed with respect to the top of the sample. The X-ray beam was centered 20 $\mu$m below the sample surface during the experiment (shaded green), while the X-ray beam was centered 20 $\mu$m, 40 $\mu$m, and 60 $\mu$m below the sample surface for various simulations used in Gaussian Process Regression model training.}
      \label{fig:exp_setup}
\end{figure}

The wall specimen was then used for an \emph{in situ} laser melting experiment at Sector 1-ID of the Advanced Photon Source. An existing \emph{in operando} laser powder bed fusion (LPBF) simulator \cite{Zhao2017} consisting of a laser and control, sealed chamber, and sample staging was used for laser melting. A schematic of the experimental geometry is provided in Fig. \ref{fig:exp_setup}. The laser system was composed of a ytterbium fiber laser (IPG YLR-500-AC) controlled with an intelliSCAN$_{de}$ 30. Using this system, the laser was rastered along the top of the wall specimen in the $\bm{x}$ direction at velocity $\bm{v}$ with magnitude of 0.05 m/s and power $P$ of 120 W (2,400 J/m), generating a relatively large high-temperature region and resulting residual stresses. During laser melting, a primarily uniaxial tensile residual stress developed along the top of the wall specimen. This process consisted of i) rapid expansion during heating, ii) compressive stresses sufficiently high to generate plastic flow and permanent compression, and iii) extension and tensile loading as the specimen cooled to accommodate compatibility. The primary tensile mechanical ($\varepsilon^E_{xx}$) strains that develop during cooling are also illustrated on the specimen in Fig \ref{fig:exp_setup}.

\begin{figure}[h]
      \centering \includegraphics[width=0.8\textwidth]{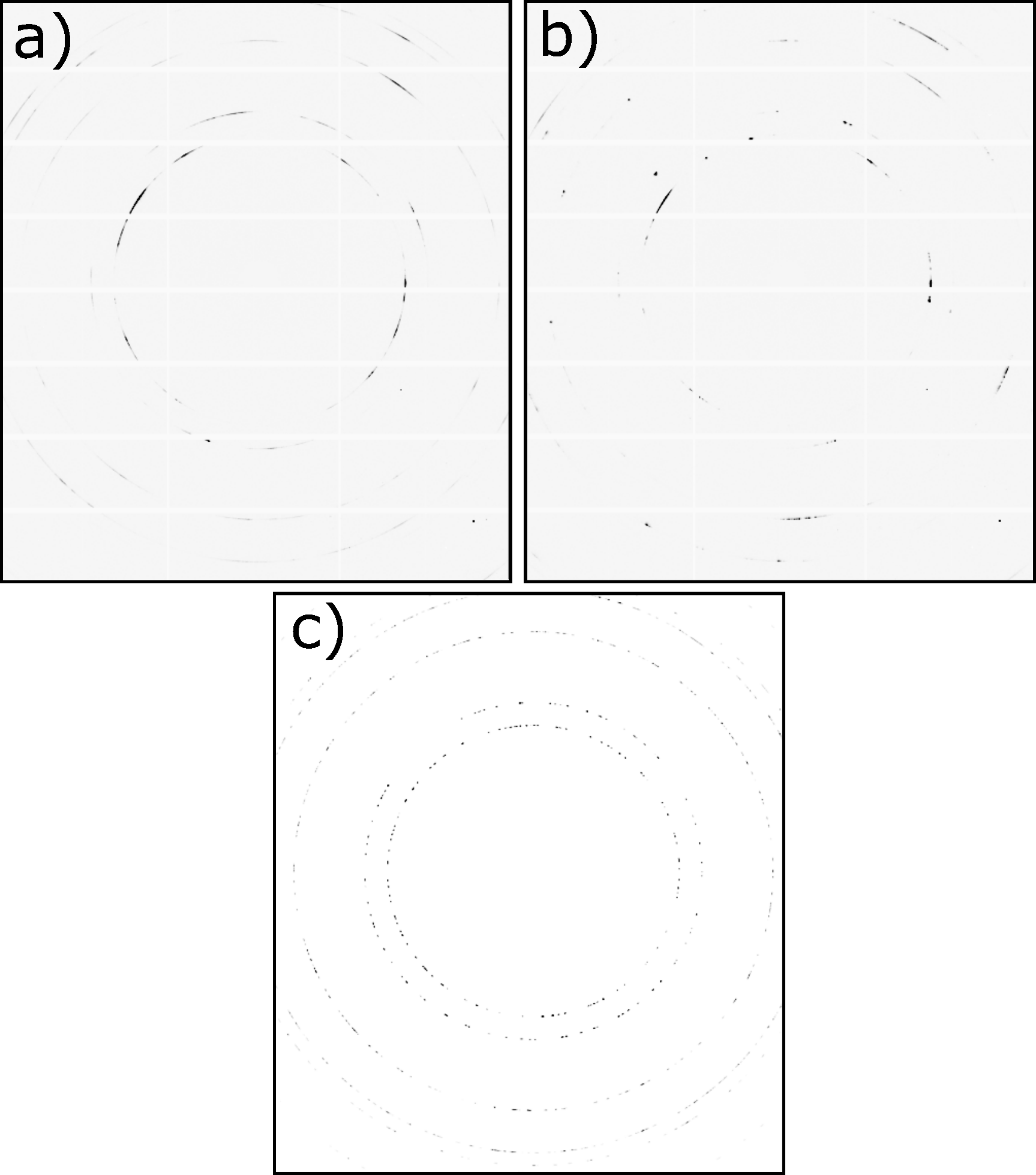}
      %\vspace{-0.15in}
      \caption{Representative diffraction images from a) the experimental sample before laser melting, b) the experimental sample after laser melting, c) a thermomechanical simulation used for Gaussian Process Regression surrogate model training. Maximum intensity thresholds have been selected for each image to make the diffraction peaks and their extent visible.}
      \label{fig:demo}
\end{figure}

During the laser melting, the wall specimen was illuminated by a 61.332 keV X-ray beam (Yb K$\alpha$) with dimensions of 50 $\mu$m (along $\bm{x}$) $\times$ 30 $\mu$m (along $\bm{z}$). The energy divergence of the beam was $5 \times 10^{4}$. The diffraction data were collected using a Pilatus3 X CdTe 2M detector sitting 752 mm downstream of the sample. The beam center was placed 20 $\mu$m below the top of the specimen and the sample and beam were fixed as the laser passed over the diffraction volume. The detector was positioned such that five complete (111, 200, 220, 311, 222) and 1 nearly-complete (400) diffraction peaks were measured on the detector (maximum 2$\theta$ angle of 13$^\circ$). Diffraction data were collected for 4 s at a rate of 250 Hz and exposure time of 1 ms, with data collection synchronized such that the laser passed over diffraction volume at 0.2 ms. Representative diffraction images collected  before and after laser melting are shown in Figs. \ref{fig:demo}a and \ref{fig:demo}b respectively. We note that with the rapid data collection rate, minor phases such as $\delta$ phase do not diffract sufficient intensity to be characterized.

\begin{figure}[h]
      \centering \includegraphics[width=0.9\textwidth]{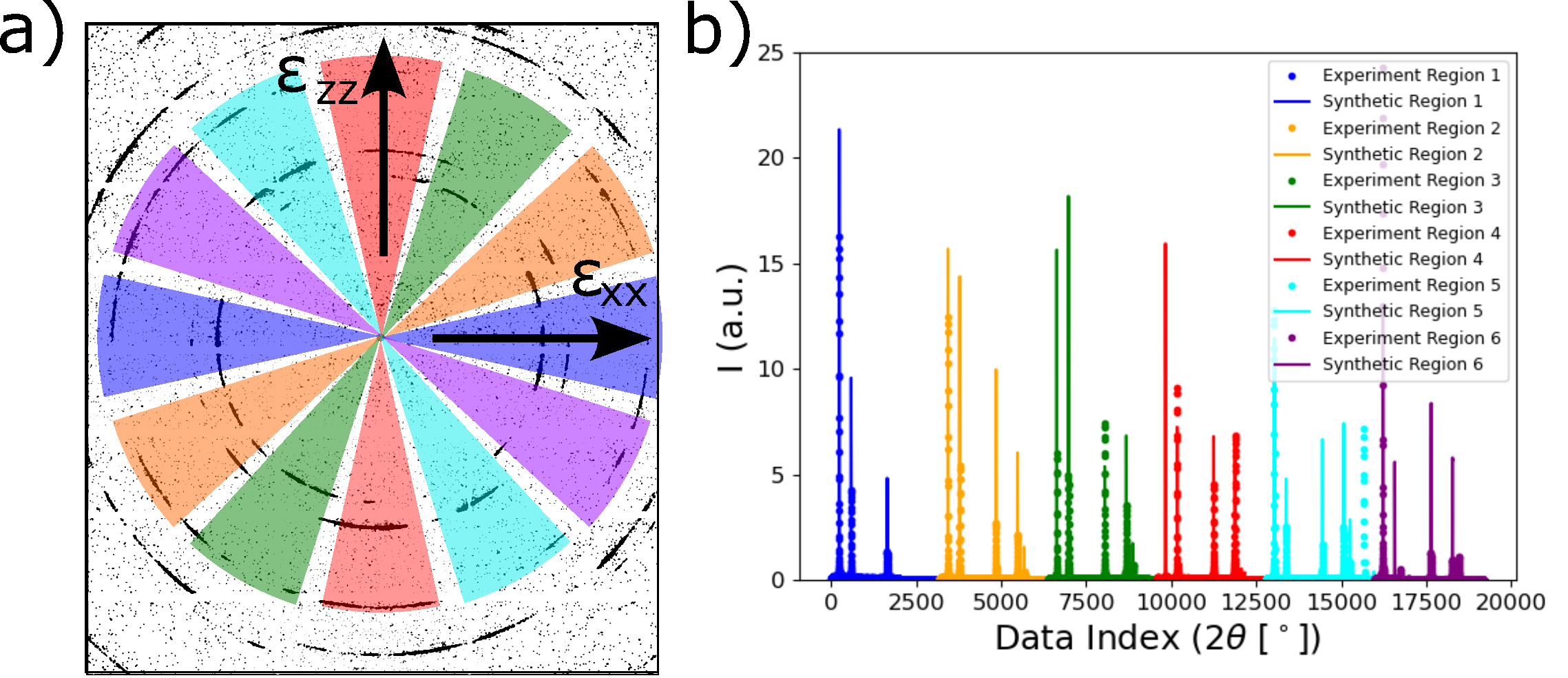}
      %\vspace{-0.15in}
      \caption{a) Illustration of the six sets of azimuthal bins (colored) around which diffraction images were integrated. b) Concatenation of the 1D intensity data from the different bins for use in ML models. A comparison of representative experimental and simulated diffraction data are shown.}
      \label{fig:bins}
\end{figure}

For use in ML models (described below), each diffraction image was then integrated into six different azimuthal bins which are illustrated in Fig. \ref{fig:bins}a. The bins (blue) along the image horizontal are aligned with projections of the strain tensor $\bm{\varepsilon}$ near the $\bm{x}$ direction ($\varepsilon_{xx}$), while vertical bins (red) are projections near the $\bm{z}$ direction ($\varepsilon_{zz}$). The other four sets of bins (orange, green, teal, and purple) further capture the anisotropy of the diffraction ring evolution. Once integrated, the six sets of 1D diffraction profiles were concatenated into a single vector illustrated in \ref{fig:bins}b. This choice of binning strategy was made as a compromise between probing a sufficient number of grains along different directions while still capturing the anisotropy of diffraction ring shape changes due to multiaxial elastic strains.

\section{Training Data Generation} % GPR Training}
\label{sec:tdata}

\subsection{Heat Transfer and Fluid Flow}

The heat transfer and fluid flow modeling that is used to generate input temperature fields is discussed in detail in Mukherjee et al. \cite{Mukherjee2018,Mukherjee2018a}. The model is designed to be capable of simulating the laser powder bed fusion process, but modeling of melting of solid materials is also possible. The code uses a finite difference scheme to simultaneously solve for conservation of mass, energy, and momentum. An adaptive grid is used with the calculation grid refining around the current position of the laser spot and melt pool (i.e., the regions with largest thermal gradients and the melt pool interface). The conditions and material parameters mirror those found in the previous effort upon which this work builds \cite{Lim:xx5027} with the only modification being that the simulations were re-performed with an extended cool-down period (5 s) to allow the specimen to cool down closer to room temperature for more input into the thermomechanical processing. The material properties used for simulations were calculated using JMatPro and are summarized in Table \ref{tab:in625_SimProps}.

\begin{table}[h]
    \centering
    \caption{IN625 model parameters used in the heat transfer and fluid flow modeling calculated using JMatPro.  Thermal conductivity and specific heat are temperature $T$ dependent (units K).}
    \label{tab:in625_SimProps}
    \begin{tabular}{| m{8.2cm} | m{5cm} |} \hline
        \textbf{Model Parameter} & \textbf{Value}  \\
        \hline
        Density (\si{kg/m^{3}}) & 8440 \\
        Solidus Temperature (\si{K}) & 1563 \\
        Liquidus Temperature (\si{K}) & 1623 \\
        Specific Heat (\si{J/kg/K}) & $360.4+0.26~T - 4\times 10^{-6}~T^2$ \\
        Thermal Conductivity (\si{W/m/K}) & $0.56 + 2.9 × \times 10^{-2}~T - 7 \times 10^{-6}~T^2$\\
        Latent Heat of Fusion (\si{J/kg}) & 209.2 $\times 10^3$ \\
        Viscosity (\si{kg/m/s}) & 5.3 $\times 10^{-3}$ \\
        Temperature Coefficient of Surface Tension (\si{N/m/K}) & -0.37 $\times 10^{-3}$ \\
        Surface Tension (\si{N/m}) & 1.82 \\
        Absorptivity Factor & 0.3 \\
        Emissivity Factor & 0.4 \\ 
    \hline
\end{tabular}
\end{table}

A series of simulations were performed rastering a simulated laser heat source over the top of wall specimens with the same thickness (0.53 mm) and height (3 mm) as in the experiment. The length of the specimen was 11.5 mm to reduce computational cost. The heat source moved along the top surface (normal $\bm{z}$) in the $\bm{x}$ direction. Following \cite{Lim:xx5027}, nine different simulations were performed around the nominal experimental laser setting (power and speed varied) with the same approximate laser size and are summarized in Table \ref{tab:test_mat}. The extra thermomechanical simulations beyond 120 W and 0.04 m/s help to compensate for any uncertainty in the laser parameters applied during the experiment. After the heat source passed over the length of the specimen, the sample was allowed to cool for 5 s as mentioned. The primary outputs are temperature fields $T(\bm{r})$, where $\bm{r}$ denotes position, saved at a rate of 500 Hz as the heat source moved and then 100 Hz as the sample cooled for a total of 150 time steps. As mentioned, the code adaptively changes the calculation grid throughout the simulation for computational efficiency. To prepare the temperature fields for use in the next steps of the workflow (Fig. \ref{fig:overview}), the simulations were post-processed using bilinear interpolation to remap the output to a regular grid with a point spacing of 20 $\mu$m.

\begin{table}
\centering
 \caption{Table summarizing the data sets generated for Gaussian Process Regression. Thermomechanical simulations were performed with different laser powers and velocities, in addition to X-ray simulations moving the position of the X-ray beam (X-ray pos.) with respect to the top of the sample.}
 \vspace{3mm}
\begin{tabular}{|c|c|c|c|c|}
\hline
No. & Power (W)  &  Velocity (m/s) &  X-ray Pos. ($\mu$m) \\ \hline
1 & 100  &  0.04 &  20 \\
2 & 100  &  0.04 &  40 \\
3 & 100  &  0.04 &  60 \\
4 & 100  &  0.05 &  20 \\
5 & 100  &  0.05 &  40 \\
6 & 100  &  0.05 &  60 \\
7 & 100  &  0.06 &  20 \\
8 & 100  &  0.06 &  40 \\
9 & 100  &  0.06 &  60 \\
10 & 120  &  0.04 &  20 \\
11 & 120  &  0.04 &  40 \\
12 & 120  &  0.04 &  60 \\
13 & 120  &  0.05 &  20 \\
14 & 120  &  0.05 &  40 \\
15 & 120  &  0.05 &  60 \\
16 & 120  &  0.06 &  20 \\
17 & 120  &  0.06 &  40 \\
18 & 120  &  0.06 &  60 \\
19 & 140  &  0.04 &  20 \\
20 & 140  &  0.04 &  40 \\
21 & 140  &  0.04 &  60 \\
22 & 140  &  0.05 &  20\\
23 & 140  &  0.05 &  40 \\
24 & 140  &  0.05 &  60 \\
25 & 140  &  0.06 &  20 \\
26 & 140  &  0.06 &  40 \\
27 & 140  &  0.06 &  60 \\ \hline
\end{tabular}

    	 \label{tab:test_mat}
\end{table}

\subsection{Thermomechanical Model}
\label{sec:mech}

Once the 9 time series temperature fields were calculated from the heat transfer and fluid flow modeling, they are used as external data inputs for an elasto-plasticity model in ANSYS to generate thermal $\varepsilon^T(\bm{r})$ and elastic strain $\bm{\varepsilon}^E(\bm{r})$ fields. The element type used was eight node `brick' elements with three degrees of freedom (translations in $\bm{x}$, $\bm{y}$, and $\bm{z}$) at each node. Element sizes ranged from 125 to 140 $\mu$m (approximately 10,000 elements per simulations) along their edges. 

\begin{figure}[h!]
  \centering \includegraphics[width=0.6\textwidth]{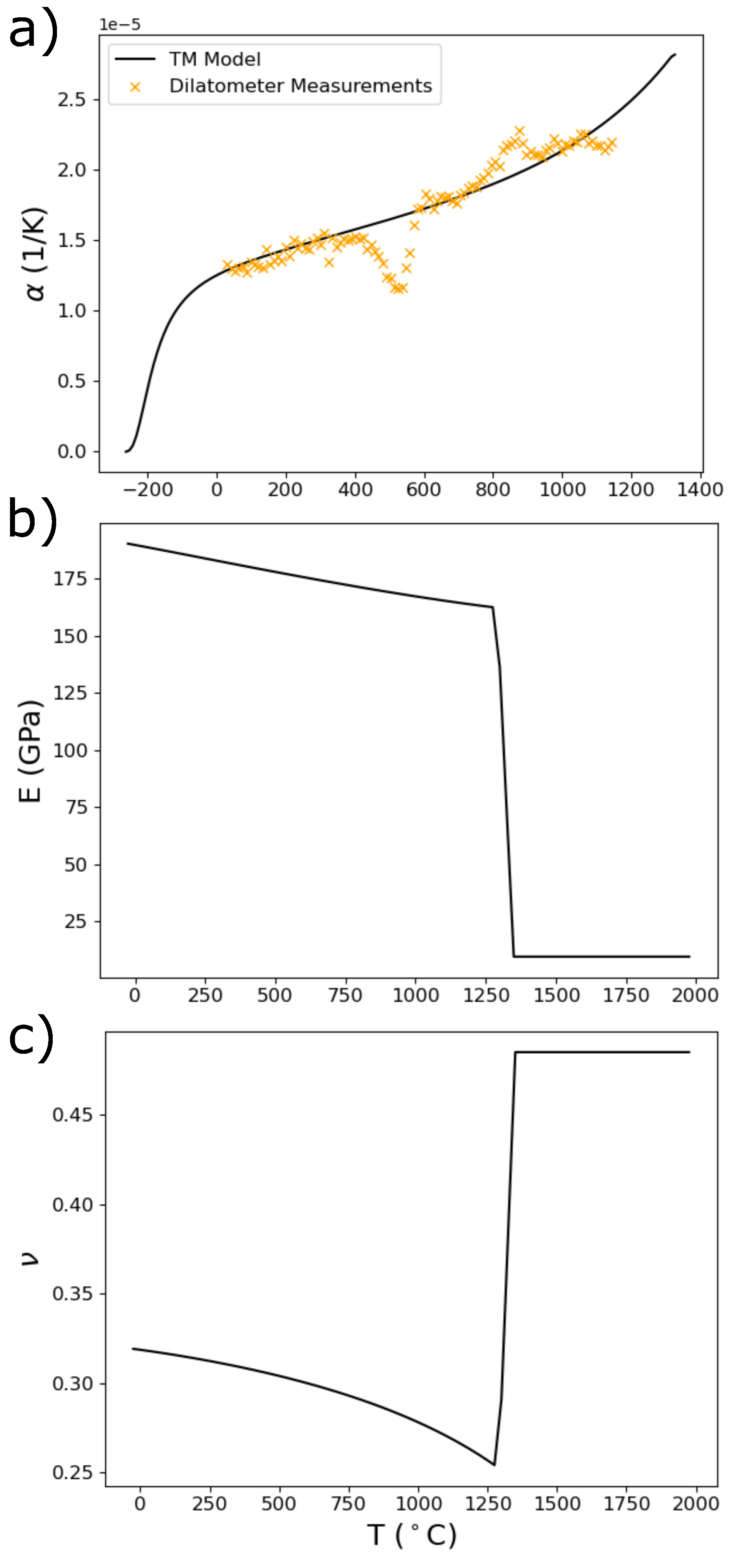}
  \caption{Temperature-dependent a) coefficient of thermal expansion $\alpha$, b) isotropic Young's modulus $E$, and c) isotropic Poisson's ratio $\nu$ used as input for the finite element simulations of the development of thermal and residual stresses in the AM IN625 wall specimens derived from the single crystal elastic moduli calculated from DFT. Dilatometer measurements of thermal expansion used to evaluate the DFT results are also provided in a).}
  \label{fig:mod_input}
\end{figure}

In the model used \cite{Taylor1970,ANSYSmechanical}, strains are additively decomposed into thermal, elastic, and plastic portions with the total deformation field being that required to maintain compatibility
\begin{equation}
    \bm{\varepsilon} = \varepsilon^T \bm{I} + \bm{\varepsilon}^E + \bm{\varepsilon}^P \quad .
\end{equation}
The thermomechanical model employed includes thermal expansion
\begin{equation}
    \varepsilon^T=\int_{T^0}^T \alpha(T) dT \quad ,
    \label{eq:thermexp}
\end{equation}
and temperature-dependent isotropic linear elasticity 
\begin{equation}
    \bm{\varepsilon}^E=\frac{1+\nu (T)}{E(T)}\bm{\sigma}-\frac{\nu (T)}{E(T)} Tr{(\bm{\sigma})}\bm{I} \quad .
    \label{eq:elasticity}
\end{equation}
Plasticity is governed by rate-independent J2 plasticity and linear hardening where yielding occurs when
\begin{equation}
    \sigma^{Y}(T)+H(T) \tilde{\varepsilon}^P=\tilde{\sigma}
    \label{eq:plasticity}
\end{equation}
where $\sigma^{Y}(T)$ is the initial temperature-dependent yield strength, $H(T)$ is the temperature-dependent hardening rate, $\tilde{\varepsilon}^P$ is the equivalent plastic strain, and $\tilde{\sigma}$ is the equivalent (von Mises) stress. If the stresses are sufficient to initiate yielding (and ultimately the development of residual stress), plastic flow occurs. As the literature is limited regarding temperature-dependent properties of specific alloy compositions, the temperature-dependent coefficient of thermal expansion $\alpha(T)$ and the elastic moduli, Young's modulus $E(T)$ and Poisson's ratio $\nu(T)$, were determined using first-principles DFT. The process by which these values were calculated are presented in Appendix A1. The thermal expansion and elastic moduli used for thermomechanical model input are provided in Fig. \ref{fig:mod_input}. Temperature-dependent coefficient of thermal expansion measured from the same IN625 build as the experimental thin wall with dilatometry and used to evaluate the DFT results are also provided in Fig. \ref{fig:mod_input}a. The tabulated yield and hardening parameters were determined from a combination of mechanical tests ($\leq 500 ^\circ$C) performed on the same material and IN625 parameters built into the ANSYS package (Table \ref{tab:yield}). 

\begin{figure}[h!]
  \centering \includegraphics[width=1.0\textwidth]{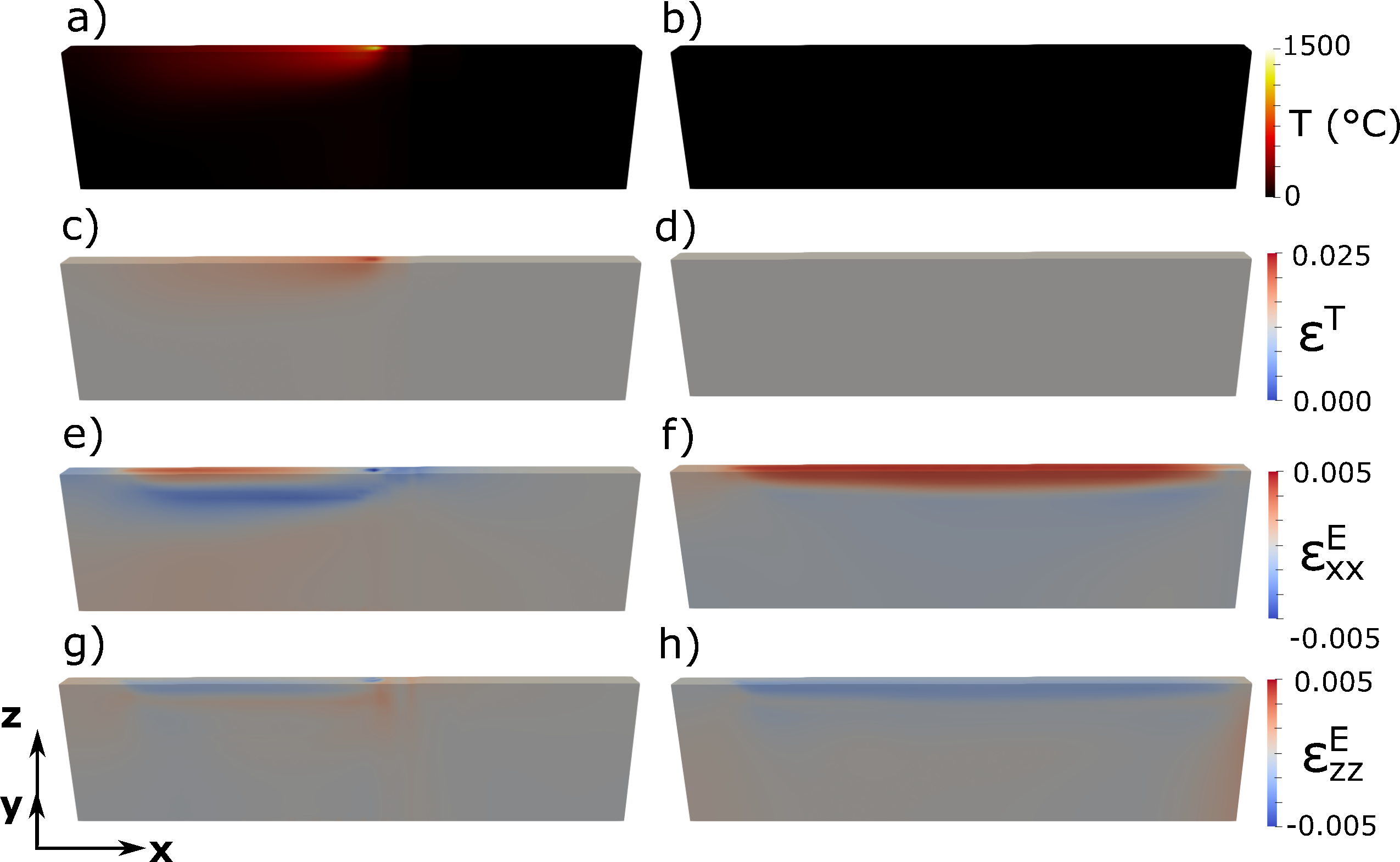}
  \caption{Representative fields from the thermomechanical simulations ($P=120$ W, $v=0.05$ m/s, and spot diameter of 100 $\mu$m) as the laser is passing overhead: a) $T$, c) $\varepsilon^T$, e) $\varepsilon^E_{xx}$, and g) $\varepsilon^E_{zz}$. Representative fields from the same laser parameter simulation after cooling: b) $T$, d) $\varepsilon^T$, f) $\varepsilon^E_{xx}$, and h) $\varepsilon^E_{zz}$.}
  \label{fig:ex_dist}
\end{figure}

Each time step takes the previously calculated thermal field and calculates local thermal strains (Eq. \ref{eq:thermexp}) from the temperature-dependent coefficient of thermal expansion previously calculated using DFT \cite{shang2024revisiting}. A strain field is then calculated which satisfies both compatibility (in conjunction with the thermal and plastic strain) and mechanical equilibrium. In all thermomechanical simulations, the thermal gradient was sufficient to initiate yielding and plastic flow. While plastic strain does not directly alter the crystal lattice and resulting diffraction, the deformation incompatibility created by the plastic strain gives rise to elastic strain and stress distributions which are measurable through diffraction. Fig. \ref{fig:ex_dist} shows representative fields ($P=120$ W, $v=0.05$ m/s, and spot diameter of 100 $\mu$m) calculated by the thermomechanical model as the laser passes over the sample (Figs. \ref{fig:ex_dist}a, \ref{fig:ex_dist}c, \ref{fig:ex_dist}e, and \ref{fig:ex_dist}g) and after cooling (Figs. \ref{fig:ex_dist}b, \ref{fig:ex_dist}d, \ref{fig:ex_dist}f, and \ref{fig:ex_dist}h). The thermal and elastic strains are plotted with different color scales due to the large differences in maximum magnitude. Important to note are the relatively large residual tensile $\varepsilon^E_{xx}$ strains that develop during the cooling process.

\begin{table}
\centering
 \caption{Temperature-dependent yield $\sigma^{Y}(T)$ and hardening $H(T)$ parameters used in the elasto-plasticity simulations.}
 \vspace{3mm}
\begin{tabular}{|c|c|c|}
\hline
Temperature ($^\circ$C) & $\sigma^{Y}(T)$ (MPa) &  $H(T)$ (MPa) \\ \hline
25  &  691 &  395 \\
500  &  615 &  395 \\
816  &  260 &  286 \\
982  &  105 &  35 \\
1093  &  57 &  20 \\
\hline
\end{tabular}
    	 \label{tab:yield}
\end{table}

To facilitate X-ray diffraction simulations from the elasto-plasticity results, a series of scripts were developed to convert ANSYS mechanical data into a format compatible with X-ray simulations. This effort included developing (i) a new  Python based binding for ASCII output of ANSYS data; (ii) a script to convert element position, stress, and strain data into the coordinate system used by HEXRD; and (iii) developing a framework to interpolate thermomechanical data on to a finer grid (20 $\mu$m spacing) necessary for diffraction simulations.

\subsection{X-ray Diffraction Simulations}

The forward modeling framework for generating synthetic diffraction patterns for model training utilizes HEXRD \cite{bernier2011far,nygren2020algorithm} and is described in more detail by \cite{Pagan2020}. Simulations build on our previous effort \cite{Lim:xx5027}, but have now been modified to include the anisotropic scattering effects of the deviatoric portion of the elastic strain tensor on the diffraction simulations. The series of nine elastic strain $\bm{\varepsilon}^E(\bm{r})$ and thermal strain fields $\varepsilon^T(\bm{r})$ (remapped to a 20 $\mu$m grid) calculated from thermomechanical simulations were used as input for the diffraction simulations. A scattering volume is built around each field point in which discrete scattering crystals can be placed. Here, grains are modeled with a 25 $\mu$m equivalent grain size so 2 crystals are inserted around each scattering volume for over 200 crystals in the diffraction volume. Each crystal is modeled with 1$^\circ$ of misorientation to provide a finite peak width perpendicular to the radial direction on the detector and ease numeric issues associated with calculating the diffraction condition. The crystals are randomly oriented which is consistent with the lack of texture found in the heat-treated IN625 being modeled. Here we project the macroscopic elastic and thermal strains onto the discrete diffracting crystals.

To model the effects of thermal and elastic strains, the reciprocal lattice vectors, $\bm{g}$, within each grain are stretched from a reference $\bm{g_0}$:
\begin{equation}
\bm{g}=(\bm{I}-\varepsilon^T\bm{I}-\bm{\varepsilon}^E) \cdot \bm{g_0} \quad.
\end{equation}
The reference lattice parameter used for calculation of $\bm{g_0}$ was 3.5981 $\AA$. At each time step, diffraction from a set of lattice planes within a grain is determined to occur based on the incoming X-ray energy and bandwidth
\begin{equation}
\bm{k}^o-\bm{k}^i=\bm{g}    \quad.
\end{equation}
where $\bm{k}^o$ and $\bm{k}^i$ are the outgoing and incoming wave vectors, respectively. The list of reciprocal lattice vectors (lattice planes) checked had a maximum 2$\theta$ of 13$^\circ$, matching the experiment. The diffraction events are then projected to the detector. Here the instrument geometry and detector were selected to match the experiment. If the temperature within a scattering volume exceeds the solidus temperature, no diffraction event is projected. Diffraction events from different sets of lattice planes are weighted by the structure factor. Drops in intensity due to absorption are neglected as this effect is minimal at high-energy. Here, we emphasize that as the reciprocal lattice vectors in grains modeled at different positions are stretched by varying amounts based on the thermal and elastic strain fields, both diffraction peak shifts \emph{and broadening} are naturally captured. Specifically we capture broadening of diffraction peak from distributions of temperature and mechanical strain in the diffraction volume, broadening from the relatively large diffraction volume size, and broadening from the finite-energy bandwidth (see \cite{Pagan2020}). We do not specifically incorporate broadening from grain to grain strain interactions from elastic anisotropy or dislocation broadening as plastic strain accumulates. In this case, the extreme thermomechanical distributions present in the diffraction volume are assumed to be the dominant source of broadening. The modeled broadening is a key feature for the ML model to learn with regards to the distributions of strain present in the diffraction volume.

Diffraction simulations for each elasto-plasticity simulation were repeated three times with the beam placed at 20 $\mu$m, 40 $\mu$m, and 60 $\mu$m from the top of the specimen and the rastering laser beam (see Fig. \ref{fig:exp_setup}). Note that the experiment was only performed with the beam centered at 20 $\mu$m from the sample surface. The purpose of this is two-fold. Performing simulations at different X-ray beam positions on the sample increased the amount of diffraction images generated for model training, and it helped to account for any uncertainty in beam placement with respect to the temperature field in the sample. In total, 27 sets of X-ray simulations in conditions similar to the experiment (3 different diffraction volumes for each of the 9 laser conditions simulated) were performed, with 75 diffraction images from each. This produced 2025 diffraction images total for GPR model training. Once the 2D diffraction patterns were simulated for the entire time series, each pattern was integrated azimuthally in the same fashion as the experimental data which is illustrated in Fig. \ref{fig:bins}. Note that the mismatch between synthetic and experimental data of the relative intensities of each peak within each color highlights the local variations of texture due to the combination of beam and grain size. After integration, background scattering was added to the same data of the same magnitude as that observed in the experiment.

\section{Machine-Learning Model and Training}
\label{sec:training}

Gaussian Process Regression \cite{Williams2006} was used for learning the mapping between diffraction data and the distributions of thermal and elastic strain present within the diffraction volume. The approach here is an extension of our previous effort of mapping isotropic diffraction ring evolution to temperature (thermal strain) fields present, to now map anisotropic diffraction ring evolution (both peak shifts and broadening) to elastic and thermal strain distributions present. GPR assumes a normal distribution of mapping functions, $f$, (here strain field metrics) from input data, $\bm{x}$, (here diffracted intensity distributions along different sample directions). The mean mapping function ($\bar{f}$) from the normal distribution is the prediction of the model. A natural benefit of the approach is that uncertainty is estimated from the variance of the mapping functions. Other ML approaches, such as neural networks, do not as readily provide measures of the uncertainty of predictions. The mapping functions learned are linear combinations of input training data $\bm{x}^*$, where the coefficients are dependent on the similarity of the training and input data. If input data are similar (here determined by Euclidean distance) to training data, those training data are more heavily weighted as dictated by the chosen covariance function (see below). In addition, if the input data for the model are not similar to any training data, the uncertainty of the model prediction increases.

For model training, we use the rational quadratic covariance function (kernel) $k$ as opposed to the more common exponentiated quadratic kernel. The rational quadratic kernel is given as
\begin{equation}
   k(\bm{x_a}, \bm{x_b}) = \left(1+\frac{||\bm{x_a}-\bm{x_b}||^2}{2 \alpha L^2}\right)^{-\alpha}, \enspace \alpha > 0
\end{equation}
where $\bm{x_a}$ and $\bm{x_b}$ are two input data points, while $\alpha$ and $L$ control the decay rate for weights. As $\alpha$ decreases, training data of increased dissimilarity from the input data are incorporated into model predictions, while $L$ provides a further control if necessary. Here, we use $L = 1$ and $\alpha = 1$. The choice of rational quadratic kernel slows the decay of the interpolation leading to more training data points being used in each prediction.

For GPR model training and testing, 26 of the data sets were used, while one (No. 13, see Table \ref{tab:test_mat}) was reserved for testing. We trained 16 different GPR models to learn mapping between the mean, maximum, minimum, and standard deviations (STDs) of $\varepsilon^T$, $\varepsilon^E_{xx}$, $\varepsilon^E_{xz}$, and $\varepsilon^E_{zz}$ within the illuminated diffraction volume and the diffraction data. For these simulations (and the experiment), $\varepsilon^E_{xz}$ is minimal across the sample, however, this still provides a further test of the model. In this diffraction geometry, the projections of $\bm{g}$ along $\bm{y}$ are minimal and as such $\varepsilon^E_{yy}$, $\varepsilon^E_{xy}$, and $\varepsilon^E_{yz}$ are not probed. We note that the trained GPR models are not fitting analytic functions to the peaks, and thus not restricted to regularly shaped diffraction peaks (e.g., Gaussian or Lorentzian) and are capable of mapping `split' peaks or those with long tails to underlying strain distributions \cite{Lim:xx5027}.

\begin{figure}[h!]
  \centering \includegraphics[width=1.0\textwidth]{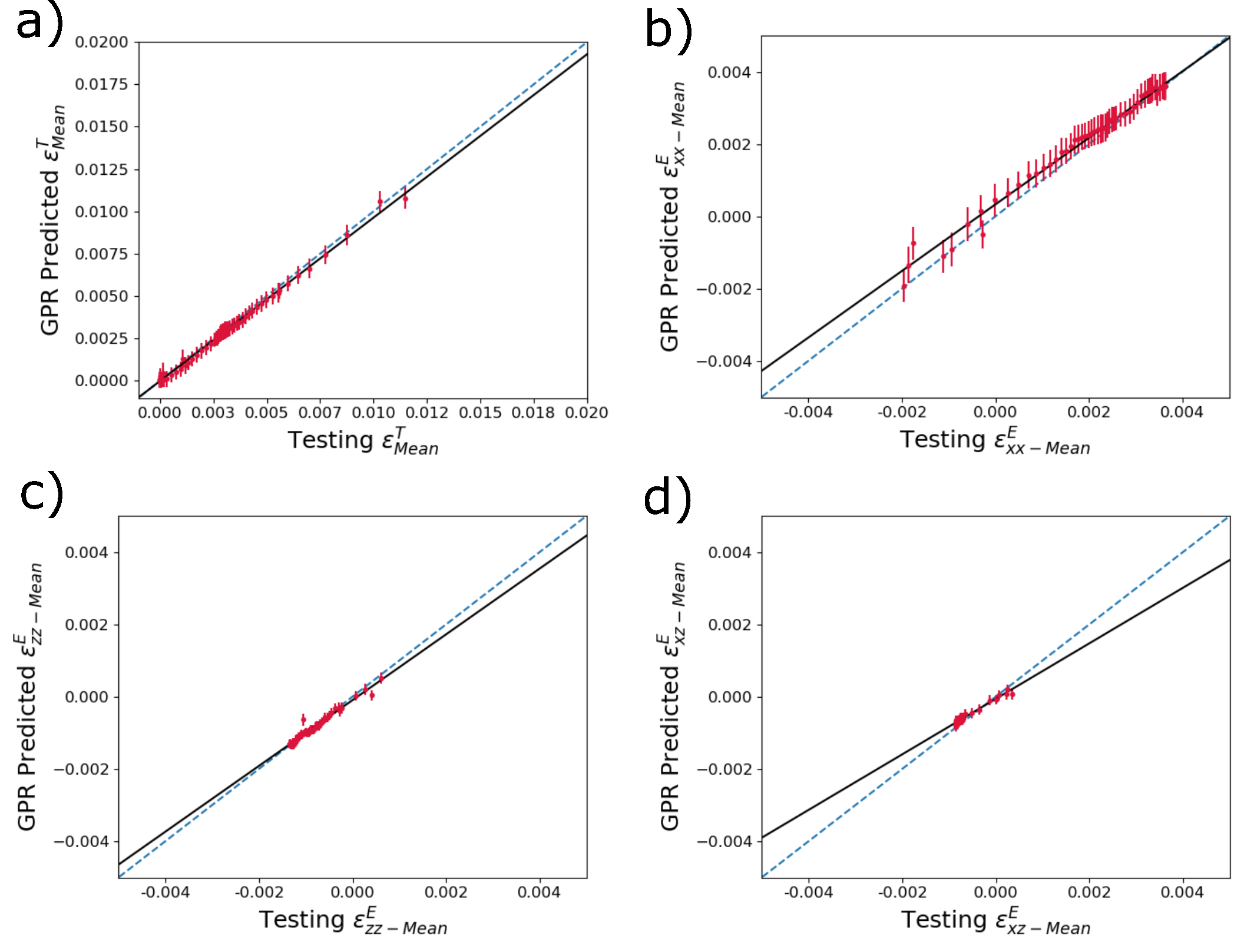}
  \caption{Accuracy predicting reserved testing data of trained surrogate models to determine the mean of a) $\varepsilon^T$, b) $\varepsilon^E_{xx}$, c) $\varepsilon^E_{zz}$, and d) $\varepsilon^E_{xz}$ within a diffraction volume from input diffraction spectra. The dashed blue line denotes one-to-one correlation, and the black line is a linear fit.
}
  \label{fig:train_mean}
\end{figure}

To examine the accuracy of the predictions of the trained GPR models, the reserved testing data set was used as input for the trained models and the output predictions were then compared to the true thermomechanical distributions quantities. Fig. \ref{fig:train_mean} shows the comparisons between mean strains in diffraction volumes from the thermomechanical model output (here serving as ground truth) compared to predictions from the trained GPR model using diffraction data as input. Figs. \ref{fig:train_mean}a, \ref{fig:train_mean}b, \ref{fig:train_mean}c, and \ref{fig:train_mean}d correspond to $\varepsilon^T$, $\varepsilon^E_{xx}$, $\varepsilon^E_{zz}$, and $\varepsilon^E_{xz}$. Note the difference in strain scales between thermal and elastic strains (due to large differences in magnitude) which will continue through the rest of the work. In general, the GPR models predictions for mean strains are accurate with the ground truth falling within the uncertainty bounds. In Fig. \ref{fig:train_mean}a, the thermal strain $\varepsilon^T_{Mean}$ predictions  tend to be closer to the ground truth at room temperature, with deviations at higher strains (temperatures). For the primary mechanical strains $\varepsilon^E_{xx-Mean}$ in \ref{fig:train_mean}b, the opposite is true, the accuracy tends to be better at larger strains. The predictions of the mean $\varepsilon^E_{zz-Mean}$ and $\varepsilon^E_{xz-Mean}$ generally fit well within the strains bounds where there is data (besides a small number of outliers), but we note that the magnitude of mean strains is relatively low for these strain components.

\begin{figure}[h!]
  \centering \includegraphics[width=1.0\textwidth]{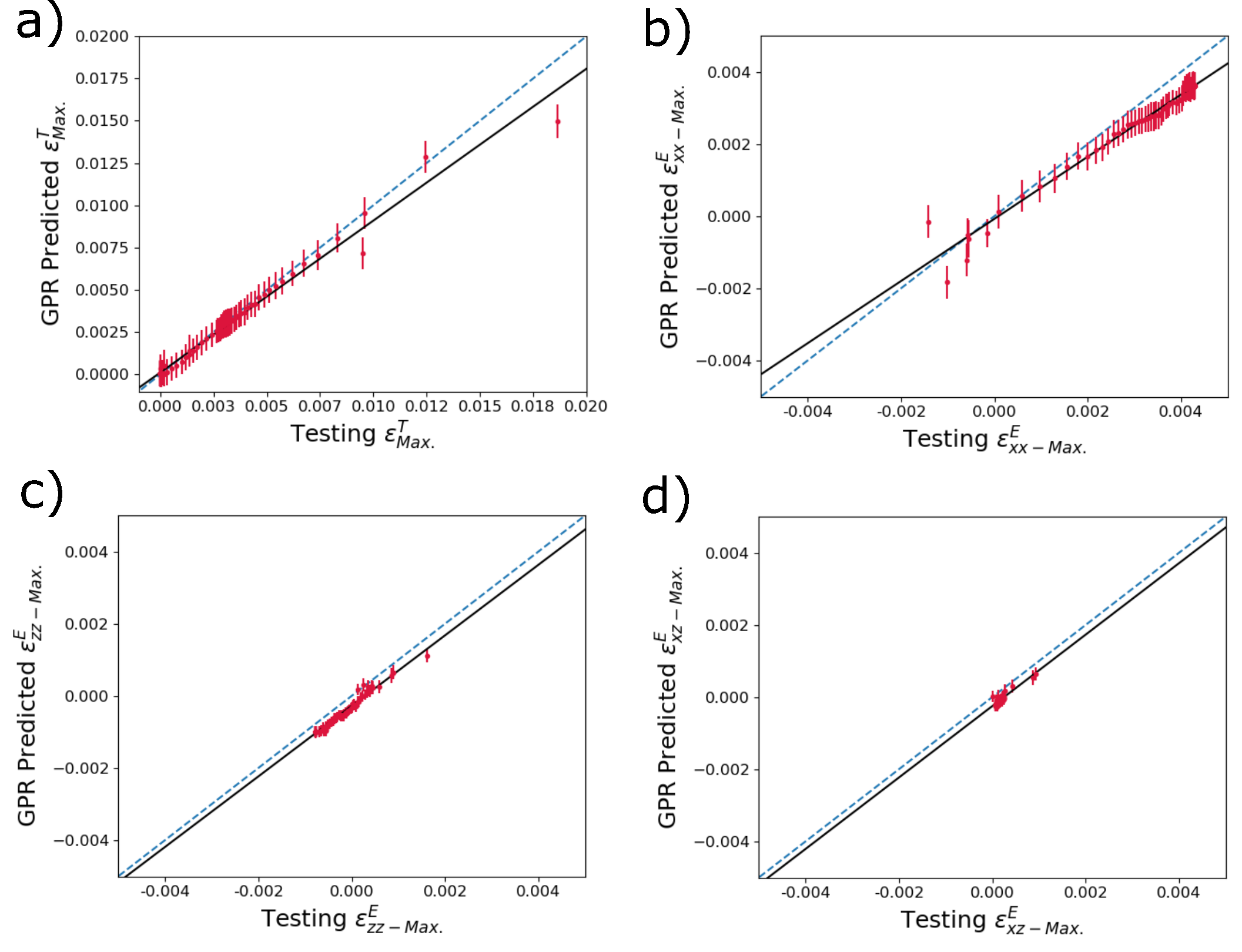}
  \caption{Accuracy predicting reserved testing data of trained surrogate models to determine the maximum of a) $\varepsilon^T$, b) $\varepsilon^E_{xx}$, c) $\varepsilon^E_{zz}$, and d) $\varepsilon^E_{xz}$  within a diffraction volume from input diffraction spectra. The dashed blue line denotes one-to-one correlation, and the black line is a linear fit.}
  \label{fig:train_max}
\end{figure}

\begin{figure}[h!]
  \centering \includegraphics[width=1.0\textwidth]{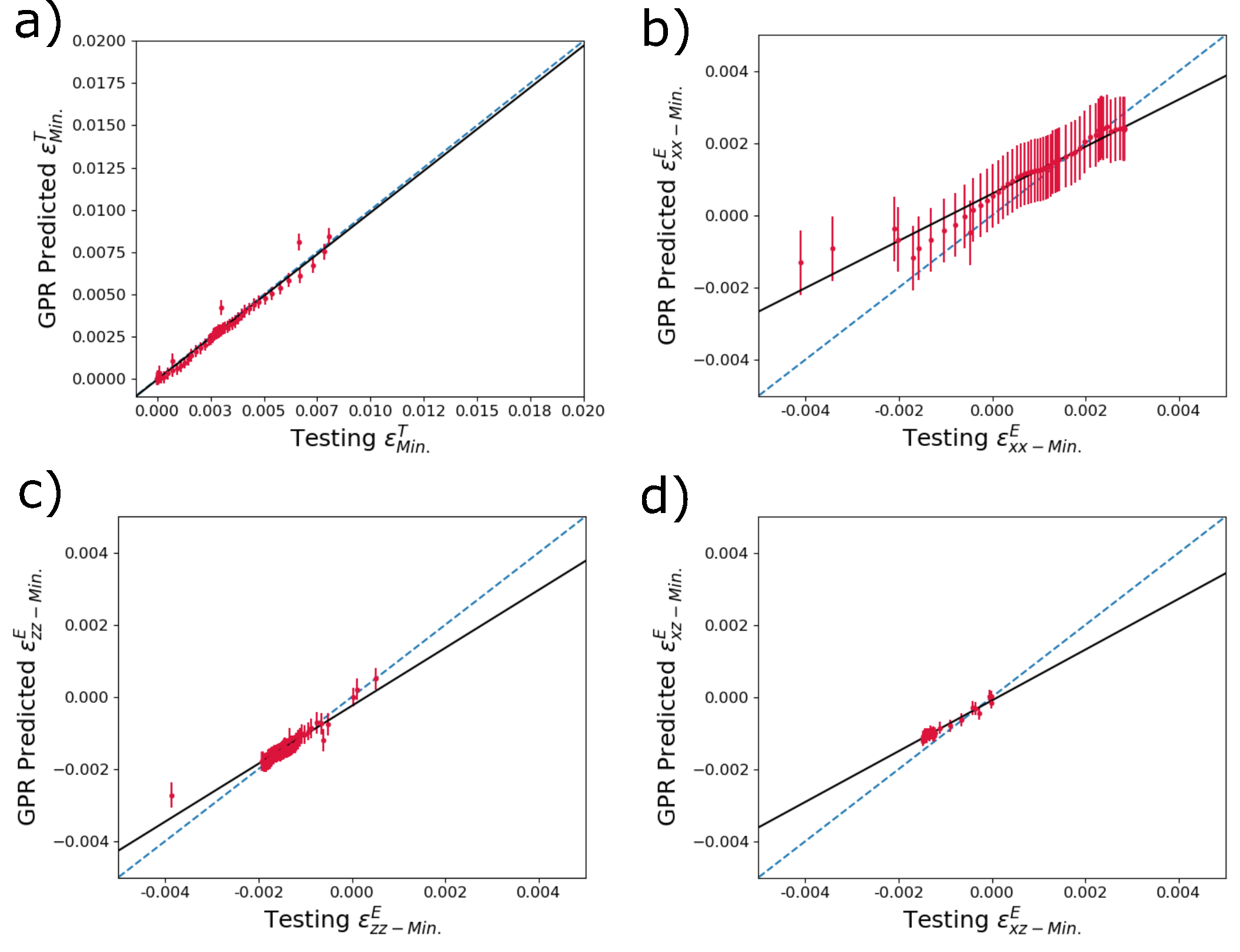}
  \caption{Accuracy predicting reserved testing data of trained surrogate models to determine the minimum of a) $\varepsilon^T$, b) $\varepsilon^E_{xx}$, c) $\varepsilon^E_{zz}$, and d) $\varepsilon^E_{xz}$ within a diffraction volume from input diffraction spectra. The dashed blue line denotes one-to-one correlation, and the black line is a linear fit.}
  \label{fig:train_min}
\end{figure}

Often of more interest are the maximum and minimum strains, particularly the maximum, within the diffraction volume. Figure \ref{fig:train_max} shows comparisons between maximum strains in the reserved testing data and the various GPR model predictions. For $\varepsilon^T_{Max.}$ in Fig. \ref{fig:train_max}a, there is a small under-prediction of the maximum strain at higher strain (temperature values). This is similar to $\varepsilon^E_{xx-Max.}$ in Fig. \ref{fig:train_max}b which also show some degree of under-prediction at higher strain values. In Figs. \ref{fig:train_max}c and \ref{fig:train_max}d, it can be seen that there is a small under-prediction of $\varepsilon^E_{zz-Max.}$ and $\varepsilon^E_{xz-Max.}$ respectively across the diffraction ranges of strain shown. The under-predictions across all strain components may be related to very small volume fractions of material in general contributing the diffraction peaks in comparison to the mean. Similarly to Fig. \ref{fig:train_max}, Fig. \ref{fig:train_min} shows comparisons between minimum strains in the reserved testing data and the various GPR model predictions. The predictions of the minimum thermal strains $\varepsilon^T_{Min.}$ in Fig. \ref{fig:train_min}a generally match well to the ground truth across the total strain range. For the minimum $\varepsilon^E_{xx}$ strains, the trends differ from predictions of the maximum in that there is a small over-prediction across the strain range. For $\varepsilon^E_{zz-Min.}$ and $\varepsilon^E_{xz-Min.}$, the strains are generally slightly under-predicted at high strain values and over-predicted at lower strain values as seen Figs. \ref{fig:train_min}c and \ref{fig:train_min}d.

\begin{figure}[h!]
  \centering \includegraphics[width=1.0\textwidth]{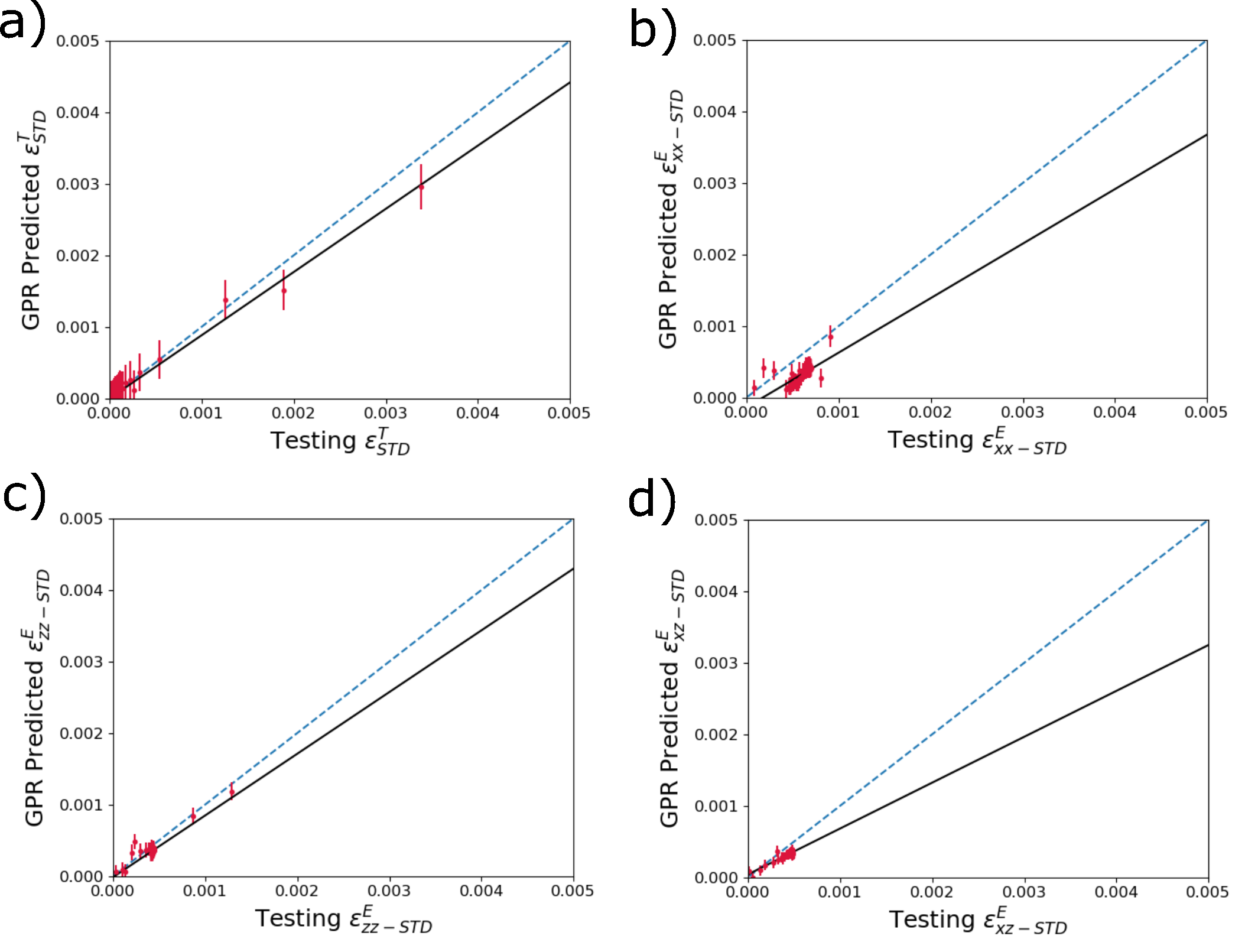}
  \caption{Accuracy predicting reserved testing data of trained surrogate models to determine the standard deviation of a) $\varepsilon^T$, b) $\varepsilon^E_{xx}$, c) $\varepsilon^E_{zz}$, and d) $\varepsilon^E_{xz}$ within a diffraction volume from input diffraction spectra. The dashed blue line denotes one-to-one correlation, and the black line is a linear fit.}
  \label{fig:train_std}
\end{figure}

The final strain metric for which GPR models were trained to extract from the diffraction data was the standard deviation (STD) of the distributions of strain within a diffraction volume. 
The results of the trained models for STD for $\varepsilon^T$, $\varepsilon^E_{xx}$, $\varepsilon^E_{zz}$, and $\varepsilon^E_{xz}$ are shown in Figs. \ref{fig:train_std}a, \ref{fig:train_std}b, \ref{fig:train_std}c, and \ref{fig:train_std}d respectively. The predictions of STD across GPR models across all strain types and components tend to under-predict the spread of the temperature distributions. This again is likely due to the very small contributions of extreme values (particularly maximums) which contribute very little intensity to the diffraction pattern. However, the GPR models are not predicting any aphysical standard deviation values (such as a negative value).

\section{Application to Experimental Data}
\label{sec:results}

The trained GPR surrogate models were used to analyze the evolution of the thermal and elastic strains within the diffraction volume during the \emph{in situ} laser melting experiment described in \S \ref{sec:expdesc}. As mentioned, the raw experimental diffraction images were azimuthally binned into six different 1D line profiles that were concatenated into a single vector (Fig. \ref{fig:bins}). Results from applying the model to a single laser pass are presented, but applying the trained surrogate models to a second laser pass at a different position on the same sample are provided in Appendix A2. Figure \ref{fig:exp_et} shows the output metrics associated with thermal strains in the diffraction volume during the \emph{in situ} experiment. The mean of the GPR predictions from each diffraction measurement are shown with black dots, while the red error bars are the variance of the GPR predictions which can be used as a measure of uncertainty. The mean (Fig. \ref{fig:exp_et}a), maximum (Fig. \ref{fig:exp_et}b), and minimum (Fig. \ref{fig:exp_et}c) of distribution of thermal strains all show a spike in thermal strain as the laser passes over the diffraction volume at 0.2 s. In addition, the standard deviations of the distributions (Fig. \ref{fig:exp_et}d) also rapidly increase as the laser passes over the specimen and then decreases back to near zero which is expected as the entire specimen returns back to room temperature.

\begin{figure}[h!]
  \centering \includegraphics[width=1.0\textwidth]{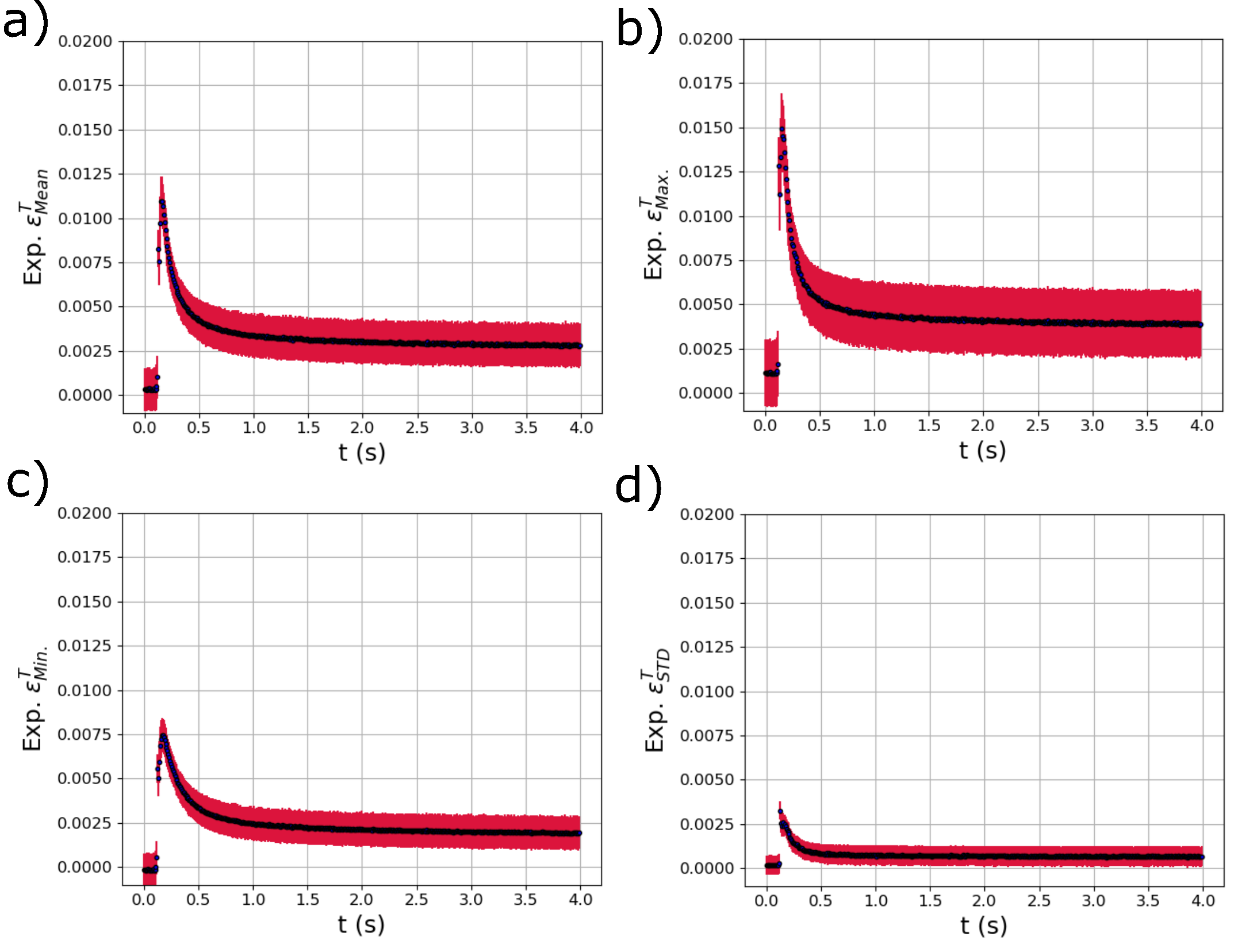}
  \caption{The evolving a) mean $\varepsilon^T_{Mean}$ b) maximum $\varepsilon^T_{Max.}$, c) minimum $\varepsilon^T_{Min.}$, and d) standard deviation $\varepsilon^T_{STD}$  of the distribution of thermal strain within the experimental X-ray diffraction volume with respect to time $t$ extracted using trained GPR surrogate models. The red error bars correspond to the square root of the variance (standard deviation) of the GPR surrogate model predictions.}
  \label{fig:exp_et}
\end{figure}

\begin{figure}[h!]
  \centering \includegraphics[width=1.0\textwidth]{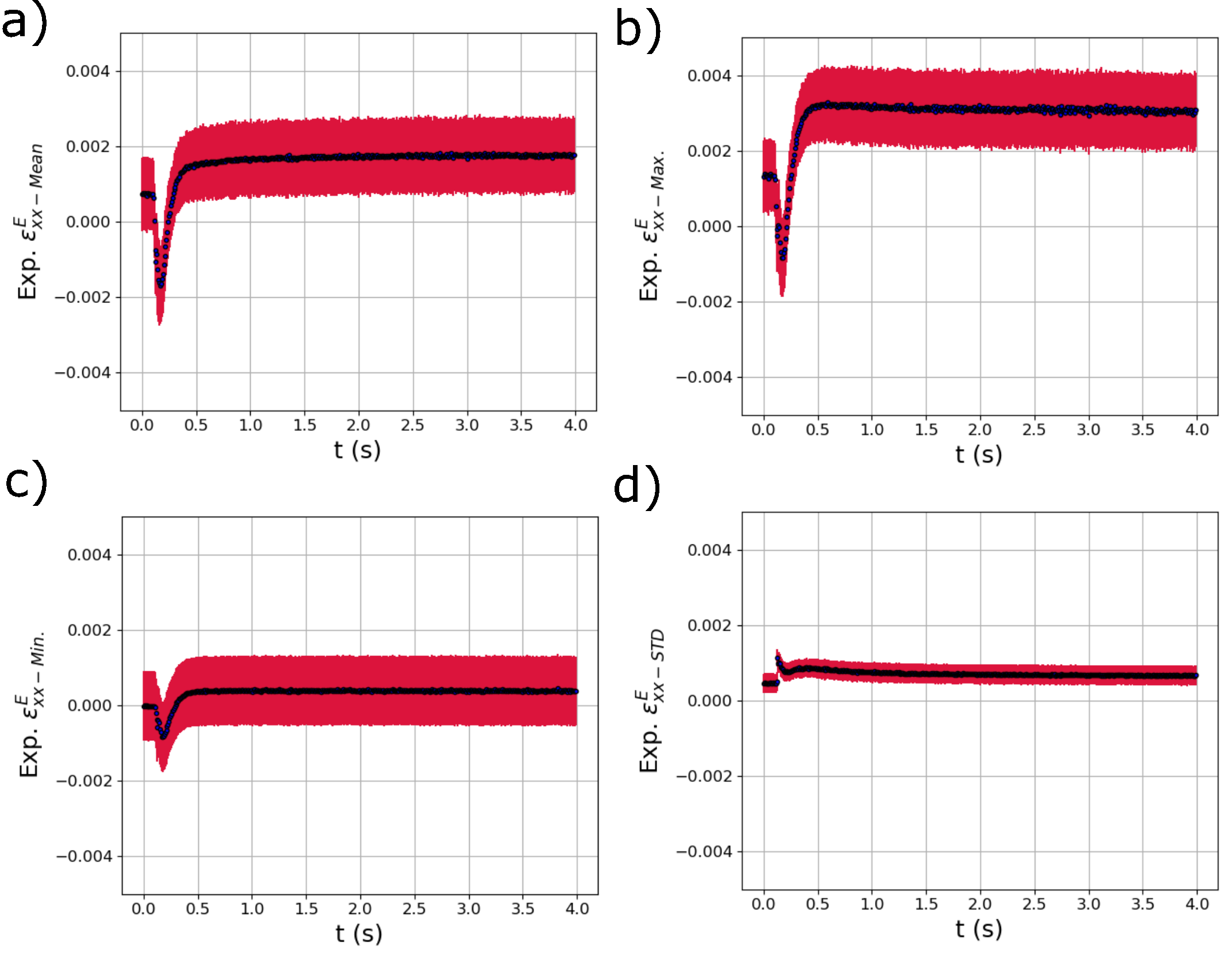}
  \caption{The evolving a) mean $\varepsilon^E_{xx-Mean}$ b) maximum $\varepsilon^E_{xx-Max.}$, c) minimum $\varepsilon^E_{xx-Min.}$, and d) standard deviation $\varepsilon^E_{xx-STD}$  of the distribution of elastic strain within the experimental X-ray diffraction volume with respect to time $t$ extracted using trained GPR surrogate models. The red error bars correspond to the square root of the variance (standard deviation) of the GPR surrogate model predictions.}
  \label{fig:exp_exx}
\end{figure}

Figure \ref{fig:exp_exx} shows the output distribution metrics from the trained GPR models for $\varepsilon^E_{xx}$. In the $\varepsilon^E_{xx}$ output metrics, there appears to be a small non-zero tensile strain at the beginning of the test (possibly from sample mounting or manufacture) which then becomes compressive as the laser passes over the sample. As the sample cools, the expected tensile residual stress develops with a larger magnitude than the original tensile strain at the beginning of the test. The general behavior of these strain distribution metrics output from the GPR model is consistent with the evolution of strains expected from simulation (see Fig. \ref{fig:ex_dist}). The most critical observation is that the GPR surrogate models appear to be able to isolate the relatively small (in comparison to the thermal strains) compressive mechanical strains that occur due to localized heating, while simultaneously capturing the large thermal strains (Fig. \ref{fig:exp_et}a). In addition, the standard deviation of $\varepsilon^E_{xx}$ shown in Fig. \ref{fig:exp_et}d increased from the start of the test and did not decay as the temperature fell, reflecting the distributions of residual strain in the specimens.

\begin{figure}[h!]
  \centering \includegraphics[width=1.0\textwidth]{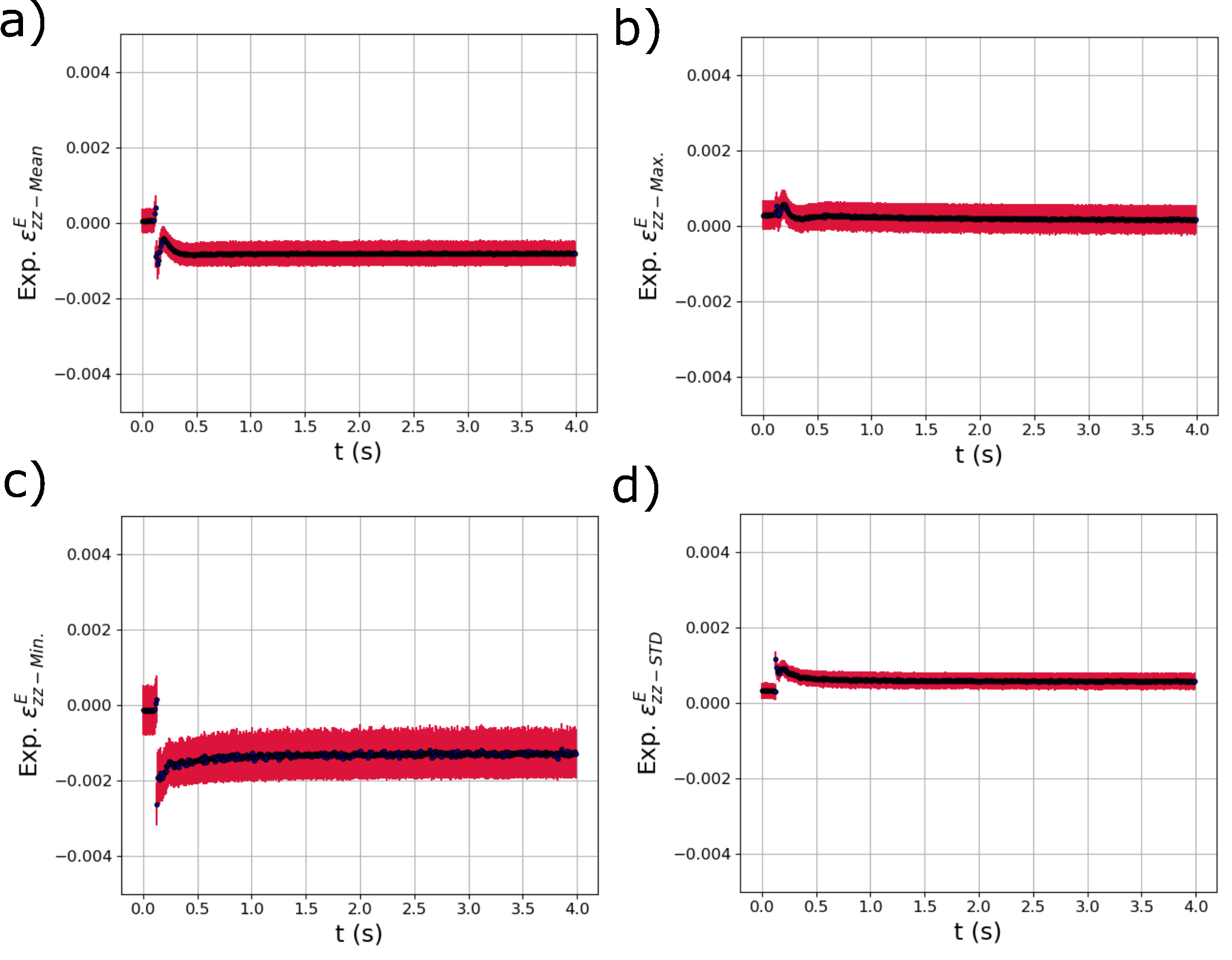}
  \caption{The evolving a) mean $\varepsilon^E_{zz-Mean}$ b) maximum $\varepsilon^E_{zz-Max.}$, c) minimum $\varepsilon^E_{zz-Min.}$, and d) standard deviation $\varepsilon^E_{zz-STD}$  of the distribution of elastic strain within the experimental X-ray diffraction volume with respect to time $t$ extracted using trained GPR surrogate models. The red error bars correspond to the square root of the variance (standard deviation) of the GPR surrogate model predictions.}
  \label{fig:exp_ezz}
\end{figure}

\begin{figure}[h!]
  \centering \includegraphics[width=1.0\textwidth]{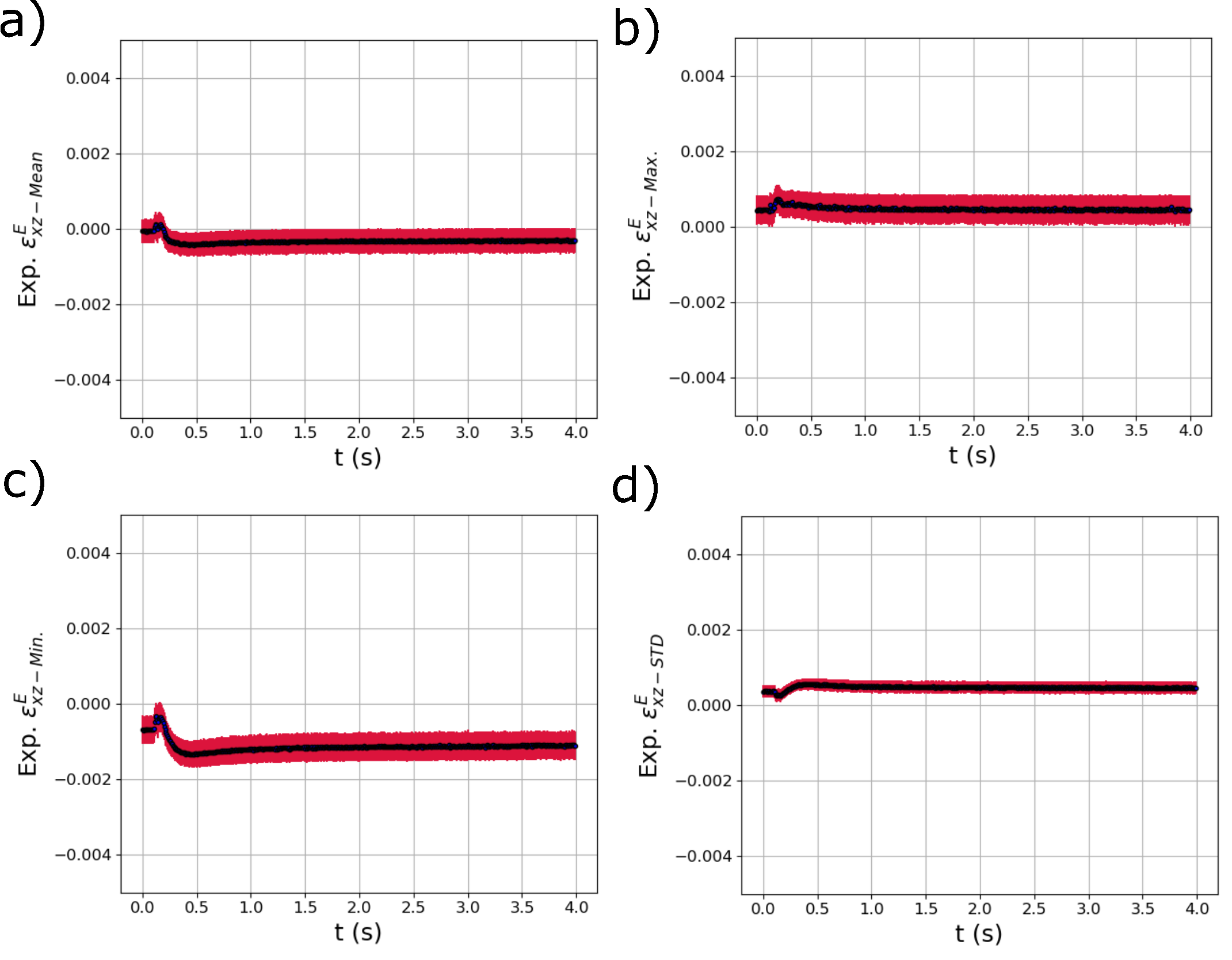}
  \caption{The evolving a) mean $\varepsilon^E_{xz-Mean}$ b) maximum $\varepsilon^E_{xz-Max.}$, c) minimum $\varepsilon^E_{xz-Min.}$, and d) standard deviation $\varepsilon^E_{xz-STD}$  of the distribution of elastic strain within the experimental X-ray diffraction volume with respect to time $t$ extracted using trained GPR surrogate models. The red error bars correspond to the square root of the variance (standard deviation) of the GPR surrogate model predictions.}
  \label{fig:exp_exz}
\end{figure}

The GPR model distribution metric outputs from the other two strain components, $\varepsilon^E_{zz}$ and $\varepsilon^E_{xz}$, are shown in Fig. \ref{fig:exp_ezz} and Fig. \ref{fig:exp_exz}. In the distributions in Fig. \ref{fig:exp_ezz}, a difference of behavior is observed between the maximum (Fig. \ref{fig:exp_ezz}b) and minimum strains (Fig. \ref{fig:exp_ezz}c) in the distribution. While the maximum strains become slightly tensile as the laser passes over the specimen, the minimum strains become negative and then remain negative through cooling. The final mean of the distribution of  $\varepsilon^E_{zz}$ is consistent with unconstrained Poisson contraction from the tensile residual strains that developed. Similar to $\varepsilon^E_{xx-STD}$, $\varepsilon^E_{zz-STD}$ peaks as the laser passes over the sample then decays to a value marginally larger than at the start of the test (Fig. \ref{fig:exp_exz}d). The distributions metrics of $\varepsilon^E_{xz}$ generally show similar trends to that of $\varepsilon^E_{zz}$ as can be see in Fig. \ref{fig:exp_exz}. However, the standard deviation of the distribution of shear strains $\varepsilon^E_{xz-STD}$ is at a minimum as the laser passes over the diffraction volume.

\section{Discussion}
\label{sec:disc}

Here, we demonstrated a machine-learning-enabled approach to analyzing complex diffraction patterns from volumes containing distributions of thermal and mechanical strains. Future application of this approach is to support the development of new classes of process diagnostics for (additive) manufacturing. Porosity from lack-of-fusion or keyholing and residual stress from unexpected process excursions are still challenges leading to failed builds. Surface optical measurements and thermography provide some insight into porosity formation, but in particular don't provide insight into residual stress development. While not immediate, rapid advances in high-brightness laboratory X-ray source technology (e.g., current metal jet \cite{larsson201124} and future thin-film diamond sources \cite{kandlakunta2019design,tan2022design}) could lead to in-chamber AM diagnostics. To characterize the thermomechanical state of the material in-chamber using these sources, non-standard analysis of diffraction data is needed. As opposed to trying to determine what the average material response is under relatively well-characterized thermomechanical loading, instead, a process diagnostic needs to quantify thermomechanical state using relatively well-understood material response (which is the approach taken here). While our approach was applied to analyzing material being laser melted, it can also be applied to any rapid, complex testing scenario where distributions of thermal and mechanical strains are expected (e.g., dynamic loading). In addition, our approach provides a path forward for analyzing other types of challenging material processing scenarios where measured lattice deformation is composed of different types of eigenstrains besides temperature (e.g., intrinsic piezoelectric or chemical).

Using the ML surrogate model trained with thermomechanical simulations for analyzing the diffraction data provides several primary benefits over more standard methods that analyze the shifts of diffraction peaks using analytical function fitting to probe temperature \cite{Hocine2020,oh2021high,oh2021microscale} or temperature and mechanical strain \cite{schmeiser2021internal,scheel2023close}. First, the lack of peak fitting allows a wider range of peak shapes to be quantitatively analyzed. For example, `split' diffraction peaks are often encountered due to the wide differences, or distributions, of temperatures between the inside and outside of a melted region. Second, the thermomechanical modeling provides a means of decoupling thermal strains and volumetric elastic strains which will deform the lattice in the same fashion. Third, the method provides metrics about the distributions of strains present which can be of significant interest for microstructural evolution as many features develop as functions of field gradients. Lastly, the GPR surrogate models provide a natural measure of uncertainty through output of the variance of the model predictions.

As was seen in \S \ref{sec:training}, the GPR models were most accurate at predicting the mean of the thermal and elastic strain distributions present in the reserved testing data. This is not surprising as the mean of the distribution also contributes to the peak of diffracted intensity.  Importantly, the GPR models appear to  be effective at fully deconvoluting thermal and elastic strains which, again, is a challenge due to the inability to directly differentiate between thermal and volumetric elastic strains in cubic materials; applying the GPR models to the \emph{in situ} experimental data yielded evolution of thermal and elastic strains that were consistent with physics-based modeling.

While still generally accurate, the predictions of maximums and minimums of the various strain distributions were subject to under- or over-predictions, usually at the extremes of the ranges of strain used for model training. The predictions of standard deviations of strain distributions were also generally under-predicted. Both of these observations are likely due to the fact that volumes of material with extreme values of strain generally do not contribute to the primary body of diffraction peaks and these extreme thermomechanical states are also extremely transient and as such have a small presence in the simulated training data. Further effort is required to determine if extreme value accuracy could be improved by increasing the number of simulations and the amount of data used for surrogate model training. In particular, increasing the number of microstructural configurations used for GPR model training may be beneficial (i.e., instantiating the same thermomechanical distributions with different grain orientation configurations).

For this effort, we took the approach of performing high-fidelity simulations for both generating thermal (finite difference heat transfer and fluid flow) and elastic (finite element elasto-plasticity) strain distributions. The rationale of this approach was that by training with high-fidelity thermal and elastic strain fields that may be present, less training data would be necessary to analyze experimental data from similar conditions. This approach appears to be successful, providing results from experimental data in line with what is expected, but the downside is that the complete modeling of the physical melting and cooling process is challenging.
To lower the modeling burden, an open question worth investigating is what fidelity of thermomechanical simulations are necessary for sufficiently accurate thermomechanical quantification from experimental measurements? Another approach may be to train the models with lower computational burden semi-analytical models \cite{weisz2020fast} or high volumes of low-fidelity simulation data. This would ease the accessibility and adoption of our proposed approach. Similarly, another avenue that requires further investigation is the sensitivity of the GPR model predictions to microstructural and material parameters used in the thermomechanical modeling for GPR training. While the GPR models are effectively connecting thermal and elastic strain distributions with diffracted intensity distributions, microstructural and material parameters will alter the space of distributions for model training, which may alter final predictions. In this work, the morphologies and size distributions of grains is not a perfect match to experiment, but the GPR model still produces rational results indicating that model accuracy may be able to be relaxed. Furthermore, here we opted for use of accurate material parameters, specifically those generated with first-principles DFT. However, this process is intensive and our presented approach would be significantly eased if literature values with less provenance or similar alloy compositions (e.g., using pure Ni properties for a superalloy) could achieve similar prediction results.

\section{Summary}
High-fidelity atomistic, heat transfer and fluid flow, thermomechanical, and X-ray scattering simulations were brought together to train machine-learning (Gaussian Process Regression, GPR) surrogate models for analyzing diffraction data with complex thermomechanical distributions of deformation present. The trained GPR models were then transferred into the experimental domain to extract thermal and elastic strain distribution metrics from \emph{in situ} diffraction data gathered during laser melting of IN625. In summary:
\begin{enumerate}
    \item The physical modeling was used to generate over 2000 diffraction images with various thermal, elastic, and microstructural configurations for GPR model training. 
    \item The GPR models are capable of separating the effects of thermal and mechanical elastic strain distributions, including at high temperatures where the magnitude of thermal strains are significantly larger.
    \item The GPR models were most accurate at predicting the mean of the thermal and elastic strain distributions from diffraction data, followed by the maximums and minimums of the distribution. In general, the GPR models tended to under-predict the standard deviation (or spread) of the thermal and elastic strains present. 
\end{enumerate}

\section*{Acknowledgments}
REL, SS, ZL, and DCP were supported by the National Institute of Standards and Technology under award 70NANB22H051. Computations for this research were performed on the Pennsylvania State University's Institute for Computational and Data Sciences' Roar supercomputer. This research used resources of the Advanced Photon Source, a U.S. Department of Energy (DOE) Office of Science user facility operated for the DOE Office of Science by Argonne National Laboratory under Contract No. DE-AC02-06CH11357.

%\appendix

\section*{Data and Code Availability Statement}

All data used for this work are available upon reasonable request. The Python-based diffraction simulation and GPR training codes are also available upon request.

\newpage
\clearpage

\bibliographystyle{elsarticle-num}
\bibliography{References.bib}

\begin{thebibliography}{10}
\expandafter\ifx\csname url\endcsname\relax
  \def\url#1{\texttt{#1}}\fi
\expandafter\ifx\csname urlprefix\endcsname\relax\def\urlprefix{URL }\fi
\expandafter\ifx\csname href\endcsname\relax
  \def\href#1#2{#2} \def\path#1{#1}\fi

\bibitem{moylan2014infrared}
S.~Moylan, E.~Whitenton, B.~Lane, J.~Slotwinski, Infrared thermography for
  laser-based powder bed fusion additive manufacturing processes, in: AIP
  Conference Proceedings, Vol. 1581, American Institute of Physics, 2014, pp.
  1191--1196.

\bibitem{everton2016review}
S.~K. Everton, M.~Hirsch, P.~Stravroulakis, R.~K. Leach, A.~T. Clare, Review of
  in-situ process monitoring and in-situ metrology for metal additive
  manufacturing, Materials \& Design 95 (2016) 431--445.

\bibitem{fox2017measurement}
J.~C. Fox, B.~M. Lane, H.~Yeung, Measurement of process dynamics through
  coaxially aligned high speed near-infrared imaging in laser powder bed fusion
  additive manufacturing, in: Thermosense: thermal infrared applications XXXIX,
  Vol. 10214, SPIE, 2017, pp. 34--50.

\bibitem{fisher2018toward}
B.~A. Fisher, B.~Lane, H.~Yeung, J.~Beuth, Toward determining melt pool quality
  metrics via coaxial monitoring in laser powder bed fusion, Manufacturing
  letters 15 (2018) 119--121.

\bibitem{montazeri2019heterogeneous}
M.~Montazeri, A.~R. Nassar, C.~B. Stutzman, P.~Rao, Heterogeneous sensor-based
  condition monitoring in directed energy deposition, Additive Manufacturing 30
  (2019) 100916.

\bibitem{dunbar2018assessment}
A.~J. Dunbar, A.~R. Nassar, Assessment of optical emission analysis for
  in-process monitoring of powder bed fusion additive manufacturing, Virtual
  and Physical Prototyping 13~(1) (2018) 14--19.

\bibitem{forien2020detecting}
J.-B. Forien, N.~P. Calta, P.~J. DePond, G.~M. Guss, T.~T. Roehling, M.~J.
  Matthews, Detecting keyhole pore defects and monitoring process signatures
  during laser powder bed fusion: A correlation between in situ pyrometry and
  ex situ x-ray radiography, Additive Manufacturing 35 (2020) 101336.

\bibitem{ashby2022thermal}
A.~Ashby, G.~Guss, R.~K. Ganeriwala, A.~A. Martin, P.~J. DePond, D.~J. Deane,
  M.~J. Matthews, C.~L. Druzgalski, Thermal history and high-speed optical
  imaging of overhang structures during laser powder bed fusion: A
  computational and experimental analysis, Additive Manufacturing 53 (2022)
  102669.

\bibitem{Kenel2016}
C.~Kenel, D.~Grolimund, J.~L. Fife, V.~A. Samson, S.~{Van Petegem}, H.~{Van
  Swygenhoven}, C.~Leinenbach, {Combined in situ synchrotron micro X-ray
  diffraction and high-speed imaging on rapidly heated and solidified Ti-48Al
  under additive manufacturing conditions}, Scripta Materialia 114 (2016)
  117--120.

\bibitem{Calta2018}
N.~P. Calta, J.~Wang, A.~M. Kiss, A.~A. Martin, P.~J. Depond, G.~M. Guss,
  V.~Thampy, A.~Y. Fong, J.~N. Weker, K.~H. Stone, et~al., An instrument for in
  situ time-resolved x-ray imaging and diffraction of laser powder bed fusion
  additive manufacturing processes, Review of Scientific Instruments 89~(5)
  (2018) 055101.

\bibitem{Cunningham2019}
R.~Cunningham, C.~Zhao, N.~Parab, C.~Kantzos, J.~Pauza, K.~Fezzaa, T.~Sun,
  A.~D. Rollett, Keyhole threshold and morphology in laser melting revealed by
  ultrahigh-speed x-ray imaging, Science 363~(6429) (2019) 849--852.

\bibitem{Hocine2020}
S.~Hocine, H.~{Van Swygenhoven}, S.~{Van Petegem}, C.~S.~T. Chang,
  T.~Maimaitiyili, G.~Tinti, D.~{Ferreira Sanchez}, D.~Grolimund, N.~Casati,
  {Operando X-ray diffraction during laser 3D printing}, Materials Today 34
  (2020) 30--40.

\bibitem{oh2021high}
S.~A. Oh, R.~E. Lim, J.~W. Aroh, A.~C. Chuang, B.~J. Gould, B.~Amin-Ahmadi,
  J.~V. Bernier, T.~Sun, P.~C. Pistorius, R.~M. Suter, et~al., High speed
  synchrotron x-ray diffraction experiments resolve microstructure and phase
  transformation in laser processed ti-6al-4v, Materials Research Letters
  9~(10) (2021) 429--436.

\bibitem{oh2021microscale}
S.~A. Oh, R.~E. Lim, J.~W. Aroh, A.~C. Chuang, B.~J. Gould, J.~V. Bernier,
  N.~Parab, T.~Sun, R.~M. Suter, A.~D. Rollett, Microscale observation via
  high-speed x-ray diffraction of alloy 718 during in situ laser melting, JOM
  73~(1) (2021) 212--222.

\bibitem{thampy2020subsurface}
V.~Thampy, A.~Y. Fong, N.~P. Calta, J.~Wang, A.~A. Martin, P.~J. Depond, A.~M.
  Kiss, G.~Guss, Q.~Xing, R.~T. Ott, et~al., Subsurface cooling rates and
  microstructural response during laser based metal additive manufacturing,
  Scientific reports 10~(1) (2020) 1--9.

\bibitem{silveira2023microstructure}
A.~C. d.~F. Silveira, R.~Fechte-Heinen, J.~Epp, Microstructure evolution during
  laser-directed energy deposition of tool steel by in situ synchrotron x-ray
  diffraction, Additive Manufacturing 63 (2023) 103408.

\bibitem{chen2023quantitative}
M.~Chen, M.~Simonelli, S.~Van~Petegem, Y.~Y. Tse, C.~S.~T. Chang, M.~G.
  Makowska, D.~F. Sanchez, H.~Moens-Van~Swygenhoven, A quantitative study of
  thermal cycling along the build direction of ti-6al-4v produced by laser
  powder bed fusion, Materials \& Design 225 (2023) 111458.

\bibitem{scheel2023close}
P.~Scheel, P.~Markovic, S.~Van~Petegem, M.~G. Makowska, R.~Wrobel, T.~Mayer,
  C.~Leinenbach, E.~Mazza, E.~Hosseini, A close look at temperature profiles
  during laser powder bed fusion using operando x-ray diffraction and finite
  element simulations, Additive Manufacturing Letters 6 (2023) 100150.

\bibitem{dass2023dendritic}
A.~Dass, C.~Tian, D.~C. Pagan, A.~Moridi, Dendritic deformation modes in
  additive manufacturing revealed by operando x-ray diffraction, Communications
  Materials 4~(1) (2023) 76.

\bibitem{Lim:xx5027}
R.~E. Lim, T.~Mukherjee, C.~Chuang, T.~Q. Phan, T.~DebRoy, D.~C. Pagan,
  {Combining synchrotron X-ray diffraction, mechanistic modeling and machine
  learning for {\it in situ} subsurface temperature quantification during laser
  melting}, Journal of Applied Crystallography 56~(4) (2023) 1131--1143.

\bibitem{levine2020outcomes}
L.~Levine, B.~Lane, J.~Heigel, K.~Migler, M.~Stoudt, T.~Phan, R.~Ricker,
  M.~Strantza, M.~Hill, F.~Zhang, et~al., Outcomes and conclusions from the
  2018 am-bench measurements, challenge problems, modeling submissions, and
  conference, Integrating Materials and Manufacturing Innovation 9 (2020)
  1--15.

\bibitem{son2020creep}
K.-T. Son, M.~E. Kassner, K.~A. Lee, The creep behavior of additively
  manufactured inconel 625, Advanced Engineering Materials 22~(1) (2020)
  1900543.

\bibitem{Zhao2017}
C.~Zhao, K.~Fezzaa, R.~W. Cunningham, H.~Wen, F.~De~Carlo, L.~Chen, A.~D.
  Rollett, T.~Sun, Real-time monitoring of laser powder bed fusion process
  using high-speed x-ray imaging and diffraction, Scientific reports 7~(1)
  (2017) 1--11.

\bibitem{Mukherjee2018}
T.~Mukherjee, H.~L. Wei, A.~De, T.~DebRoy, {Heat and fluid flow in additive
  manufacturing—Part I: Modeling of powder bed fusion}, Computational
  Materials Science 150~(April) (2018) 304--313.

\bibitem{Mukherjee2018a}
T.~Mukherjee, H.~L. Wei, A.~De, T.~DebRoy, {Heat and fluid flow in additive
  manufacturing – Part II: Powder bed fusion of stainless steel, and
  titanium, nickel and aluminum base alloys}, Computational Materials Science
  150~(April) (2018) 369--380.

\bibitem{Taylor1970}
R.~L. Taylor, K.~S. Pister, G.~L. Goudreau, {Thermomechanical analysis of
  viscoelastic solids}, International Journal for Numerical Methods in
  Engineering 2~(1) (1970) 45--59.

\bibitem{ANSYSmechanical}
Ansys, Inc., Canonsburg, PA 15317, ANSYS Mechanical APDL Structural
  AnalysisGuide (2011).

\bibitem{shang2024revisiting}
S.-L. Shang, R.~Gong, M.~C. Gao, D.~C. Pagan, Z.-K. Liu, Revisiting
  first-principles thermodynamics by quasiharmonic approach: Application to
  study thermal expansion of additively-manufactured inconel 625, Scripta
  Materialia 250 (2024) 116200.

\bibitem{bernier2011far}
J.~V. Bernier, N.~R. Barton, U.~Lienert, M.~P. Miller, Far-field high-energy
  diffraction microscopy: a tool for intergranular orientation and strain
  analysis, The Journal of Strain Analysis for Engineering Design 46~(7) (2011)
  527--547.

\bibitem{nygren2020algorithm}
K.~E. Nygren, D.~C. Pagan, J.~V. Bernier, M.~P. Miller, An algorithm for
  resolving intragranular orientation fields using coupled far-field and
  near-field high energy x-ray diffraction microscopy, Materials
  Characterization 165 (2020) 110366.

\bibitem{Pagan2020}
D.~C. Pagan, K.~K. Jones, J.~V. Bernier, T.~Q. Phan, {A Finite Energy
  Bandwidth-Based Diffraction Simulation Framework for Thermal Processing
  Applications}, Jom 72~(12) (2020) 4539--4550.

\bibitem{Williams2006}
C.~K. Williams, C.~E. Rasmussen, Gaussian processes for machine learning,
  Vol.~2, MIT press Cambridge, MA, 2006.

\bibitem{larsson201124}
D.~H. Larsson, P.~A. Takman, U.~Lundstr{\"o}m, A.~Burvall, H.~Hertz, A 24 kev
  liquid-metal-jet x-ray source for biomedical applications, Review of
  Scientific Instruments 82~(12) (2011).

\bibitem{kandlakunta2019design}
P.~Kandlakunta, A.~Thomas, Y.~Tan, R.~Khan, T.~Zhang, Design and numerical
  simulations of w-diamond transmission target for distributed x-ray sources,
  Biomedical physics \& engineering express 5~(2) (2019) 025030.

\bibitem{tan2022design}
Y.~Tan, Q.~Chen, S.~Zhou, E.~A. Henriksen, T.~Zhang, Design and optimization of
  thin-film tungsten (w)-diamond target for multi-pixel x-ray sources, Medical
  physics 49~(8) (2022) 5363--5373.

\bibitem{schmeiser2021internal}
F.~Schmeiser, E.~Krohmer, N.~Schell, E.~Uhlmann, W.~Reimers, Internal stress
  evolution and subsurface phase transformation in titanium parts manufactured
  by laser powder bed fusion—an in situ x-ray diffraction study, Advanced
  Engineering Materials 23~(11) (2021) 2001502.

\bibitem{weisz2020fast}
D.~Weisz-Patrault, Fast simulation of temperature and phase transitions in
  directed energy deposition additive manufacturing, Additive Manufacturing 31
  (2020) 100990.

\bibitem{kresse1993ab}
G.~Kresse, J.~Hafner, Ab initio molecular dynamics for liquid metals, Physical
  review B 47~(1) (1993) 558.

\bibitem{shang2010first}
S.-L. Shang, Y.~Wang, D.~Kim, Z.-K. Liu, First-principles thermodynamics from
  phonon and debye model: Application to ni and ni3al, Computational Materials
  Science 47~(4) (2010) 1040--1048.

\bibitem{zunger1990special}
A.~Zunger, S.-H. Wei, L.~Ferreira, J.~E. Bernard, Special quasirandom
  structures, Physical review letters 65~(3) (1990) 353.

\bibitem{singh2021accelerating}
R.~Singh, A.~Sharma, P.~Singh, G.~Balasubramanian, D.~D. Johnson, Accelerating
  computational modeling and design of high-entropy alloys, Nature
  Computational Science 1~(1) (2021) 54--61.

\bibitem{wang2010first}
Y.~Wang, J.~Wang, H.~Zhang, V.~Manga, S.~Shang, L.~Chen, Z.~Liu, A
  first-principles approach to finite temperature elastic constants, Journal of
  Physics: Condensed Matter 22~(22) (2010) 225404.

\bibitem{shang2012effects}
S.~Shang, D.~Kim, C.~Zacherl, Y.~Wang, Y.~Du, Z.~Liu, Effects of alloying
  elements and temperature on the elastic properties of dilute ni-base
  superalloys from first-principles calculations, Journal of Applied Physics
  112~(5) (2012).

\bibitem{shang2007first}
S.~Shang, Y.~Wang, Z.-K. Liu, First-principles elastic constants of
  $\alpha$-and $\theta$-al2o3, Applied Physics Letters 90~(10) (2007).

\bibitem{hosford1993mechanics}
W.~Hosford, The Mechanics of Crystals and Textured Polycrystals, Oxford
  engineering science series, Oxford University Press, 1993.

\end{thebibliography}

\newpage

\section*{A1. First-Principles Thermomechanical Property Calculations}

We utilized first-principles DFT using the VASP code \cite{kresse1993ab} to generate temperature-dependent IN625 elastic moduli and coefficients of thermal expansion for input into later thermomechanical modeling. Our process is described in detail in \cite{shang2010first,shang2024revisiting}, but is summarized here. In this approach, the Helmholtz energy $F$ has a 0 K static energy contribution, a vibrational contribution, and a thermal electron contribution. Here, each individual contribution to the Helmholtz energy is predicted independently, with the vibrational and thermal electron contributions calculated as a function of volume $V$ at a given temperature $T$, and then the three contributions summed. From the temperature-dependent Helmholtz energy, the equilibrium volume $V^0$ is calculated as a function of temperature. The linear coefficient of thermal expansion $\alpha$ is then calculated as
\begin{equation}
    \alpha(T)=\frac{1}{3 V^0} \left( \frac{\partial V^0}{\partial T} \right) \vert_P \quad.
\end{equation}

\begin{figure}[h!]
  \centering \includegraphics[width=0.75\textwidth]{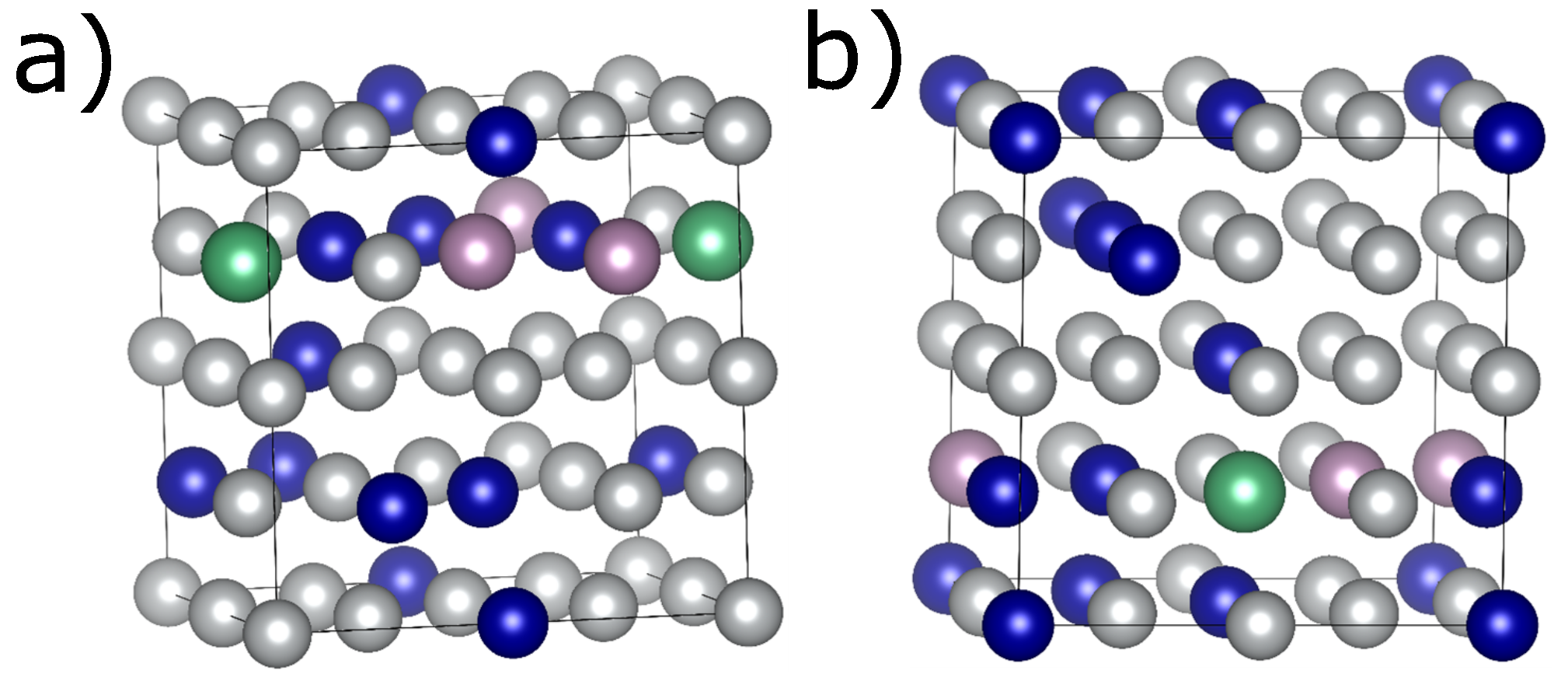}
  \caption{a) One of the generated SQSs and b) one of the generated SCRAPs of Ni$_{21}$Cr$_8$Mo$_2$Nb$_1$ (32-atom supercell), where the colored spheres represent different atom types: Ni (gray), Cr (blue), Mo (purple), and Nb (green). }
  \label{fig:cells_cal}
\end{figure}

For the IN625 property modeling, we used the compositions reported for AM Bench 2018 material \cite{levine2020outcomes} of which the same powder used was used here. In the IN625, the major alloying components besides Ni are Cr (20.61 wt\%), Nb (3.97 wt\%), and Mo (8.82 wt\%). Based on CALPHAD predictions, there are two major phases, i.e., the dominant FCC matrix ($\gamma$) and the topologically close-packed (TCP) phase, $\delta$. Here, we assume that the thermal expansion and elastic properties are primarily dependent on the FCC matrix. The composition of the FCC matrix according to CALPHAD predictions is roughly Ni$_{21}$Cr$_8$Mo$_2$Nb$_1$ \cite{shang2024revisiting}. To predict properties, relatively small cubic supercells (32-atom) were modeled, and then the results were averaged to strike a balance between accuracy and computational efficiency. We generated 6 special quasirandom structures (SQSs) \cite{zunger1990special} and 6 supercells in random approximates (SCRAPs) \cite{singh2021accelerating}. Figure \ref{fig:cells_cal} shows representative SQS and SCRAPs structures, with Fig. \ref{fig:cells_cal}a showing one of the generated SQSs and Fig. \ref{fig:cells_cal}b one of the generated SCRAPs. Temperature-dependent Helmholtz energy was calculated for each of these supercells and then used for both coefficient of thermal expansion and elastic moduli calculations, see details including first-principles settings in \cite{shang2024revisiting}. Figure \ref{fig:mod_input}a shows the calculated temperature-dependent coefficient of thermal expansion $\alpha(T)$ using this procedure. The values matched well with experimental data used in \cite{Lim:xx5027} within the temperature bounds measured.

\begin{figure}[h!]
  \centering \includegraphics[width=0.6\textwidth]{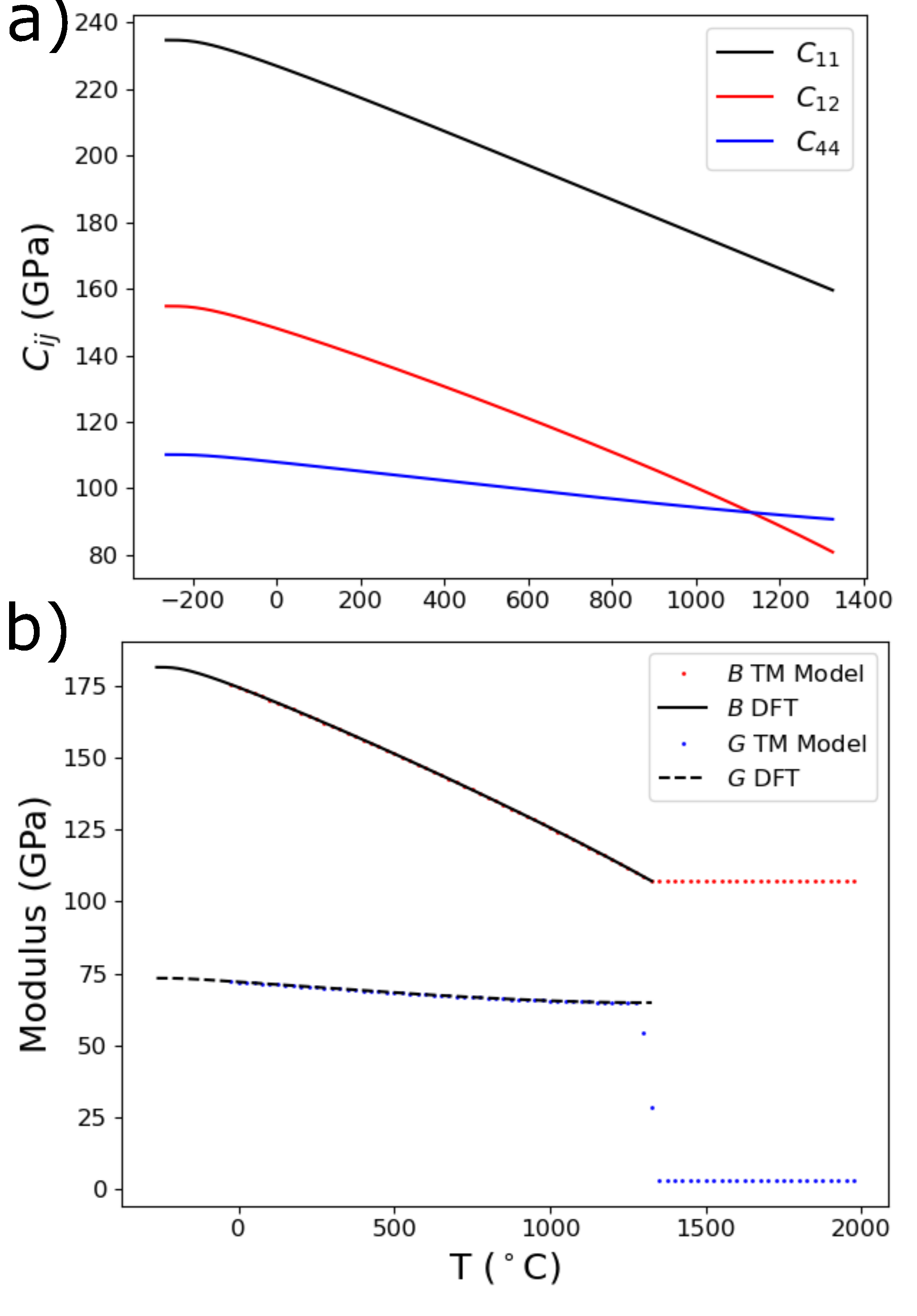}
  \caption{a) Temperature-dependent cubic single crystal elastic moduli $\bar{C}_{ij}$ of AM IN625 calculated using DFT-based quasi-harmonic analysis (QHA) and quasi-static analysis (QSA). b) Bulk modulus $B$ (black line) and shear modulus $G$ (black dashed line) calculated from the DFT single crystal elastic moduli. Also shown are the bulk modulus $B$ (red dots) and shear modulus $G$ (blue dots) used as input for the thermomechanical model (TM Model).}
  \label{fig:dft_moduli}
\end{figure}

Single crystal elastic moduli ($C_{ij}$) as a function of temperature $T$ were predicted using the quasi-static approach (QSA) \cite{wang2010first,shang2010first,shang2012effects} according to the predicted relationship between $C_{ij}$ and $V$ at 0 K and the previously calculated relationship between $T$ and $V$ (i.e., thermal expansion) \cite{shang2024revisiting}, resulting in the predicted $C_{ij}(T)$. Note that the DFT-based calculations of $C_{ij}(V)$ at 0 K were predicted using the anisotropic form of Hooke's law \cite{shang2007first}, where the employed non-zero strains applied are $\pm$ 0.01. Since both the SQSs and SCRAPs of IN625 are monoclinic configurations, averaged values of the elastic moduli were used to calculate cubic elastic moduli $\bar{C}_{ij}$, i.e., $\bar{C}_{11}=(C_{11}+C_{22}+C_{33})/3$, $\bar{C}_{12}=(C_{12}+C_{13}+C_{23})/3$, and $\bar{C}_{44}=(C_{44}+C_{55}+C_{66})/3$.  Note that the averaged $C_{ij}$ values usually have deviations less than 10\% from the mean for each monoclinic configuration, such as $C_{11}=$255.8 GPa, $C_{22}=$262.2 GPa, and $C_{33}=$271.7 GPa for a SCRAPS configuration (SCRAPS$_f$) at its equilibrium volume. Figure \ref{fig:dft_moduli}a shows the cubic single crystal moduli $\bar{C}_{11}$, $\bar{C}_{12}$, and $\bar{C}_{44}$ determined from the DFT calculations. From the figure, we see that both $\bar{C}_{11}$ and $\bar{C}_{12}$ decrease faster than $\bar{C}_{44}$ with increasing temperature.

%a) coefficient of thermal expansion $\alpha$ and b)

For input into the continuum thermomechanical model (\S \ref{sec:mech}), $\bar{C}_{11}$, $\bar{C}_{12}$, and $\bar{C}_{44}$ were used to generate isotropic elastic moduli. The mean of the Reuss and Voigt averages of an untextured cubic polycrystal \cite{hosford1993mechanics} were used to calculate Young's modulus $E(T)$ and the shear modulus $G(T)$, from which the Poisson's ratio $\nu(T)$ was calculated. For temperatures beyond the melting temperature of the material, the bulk modulus $B(T)$, of the material was assumed to remain constant while the shear modulus was reduced to 5\% of the value prior to melting. The resulting bulk modulus and shear modulus calculated from this process are provided in Fig. \ref{fig:dft_moduli}b.

\section*{A2. Duplicate Experimental Analysis}

\begin{figure}[h!]
  \centering \includegraphics[width=1.0\textwidth]{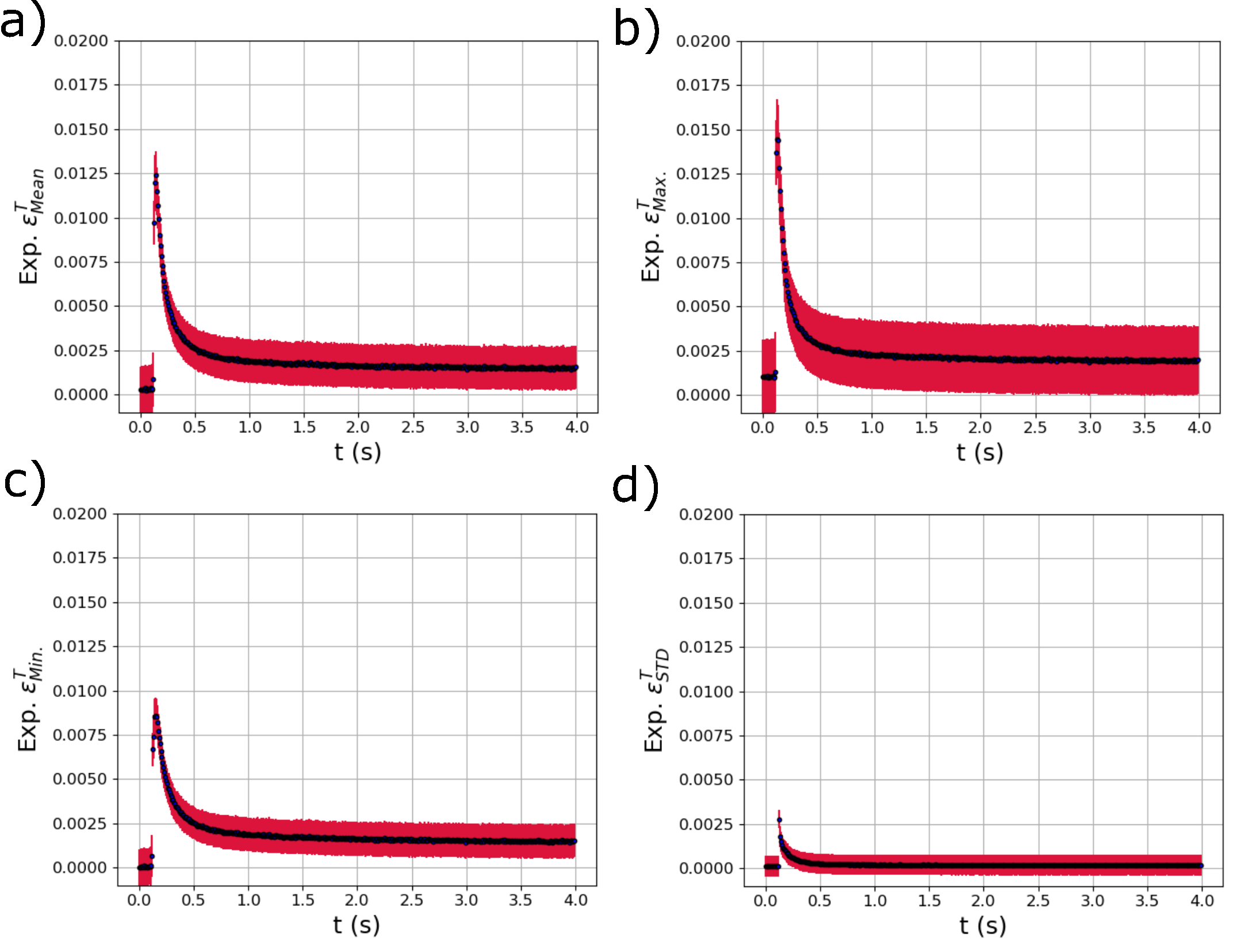}
  \caption{The evolving a) mean $\varepsilon^T_{Mean}$ b) maximum $\varepsilon^T_{Max.}$, c) minimum $\varepsilon^T_{Min.}$, and d) standard deviation $\varepsilon^T_{STD}$  of the distribution of thermal strain within the \textbf{duplicate} experimental X-ray diffraction volume with respect to time $t$ extracted using trained GPR surrogate models. The red error bars correspond to the square root of the variance (standard deviation) of the GPR surrogate model predictions.}
  \label{fig:dup_et}
\end{figure}

\begin{figure}[h!]
  \centering \includegraphics[width=1.0\textwidth]{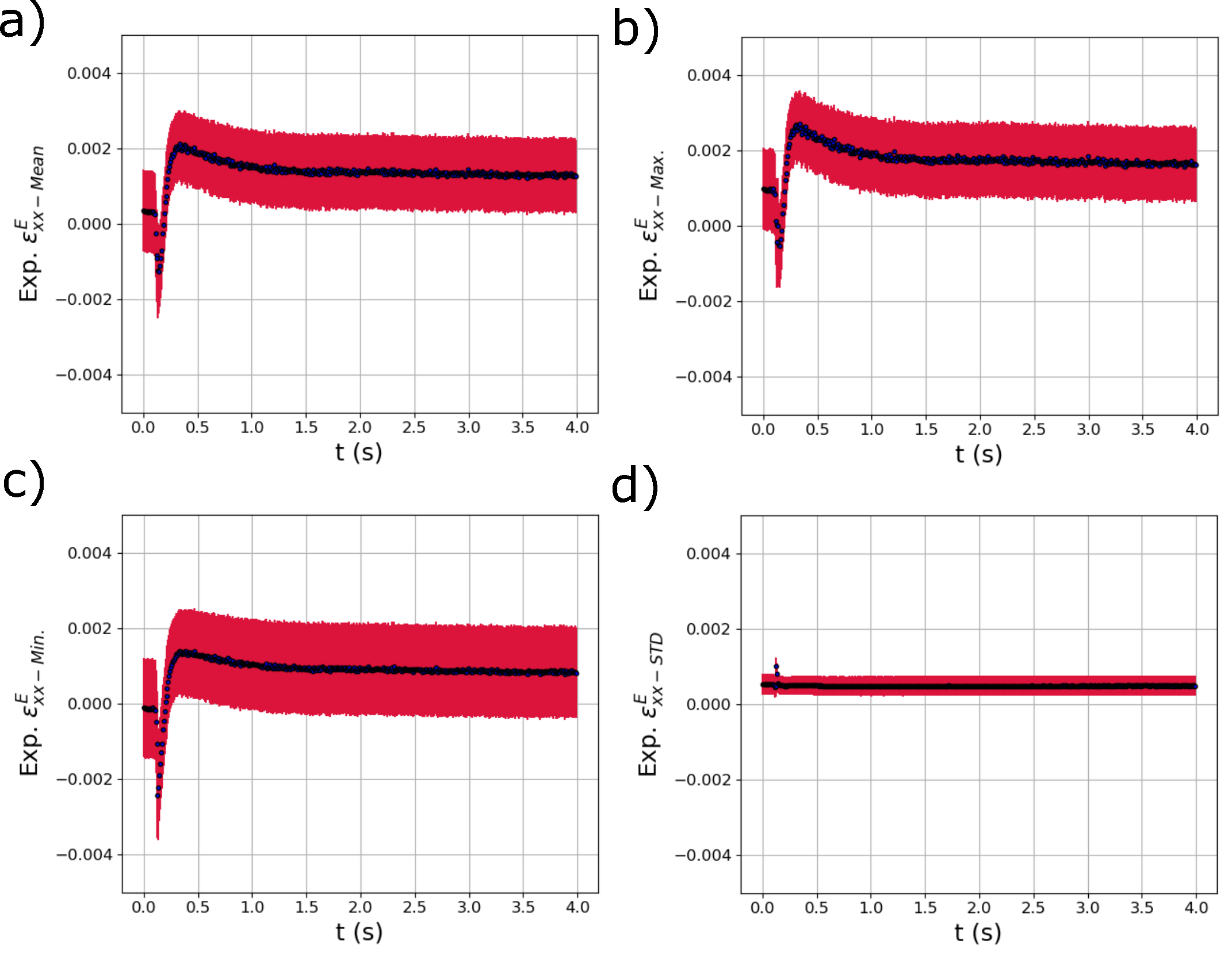}
  \caption{The evolving a) mean $\varepsilon^E_{xx-Mean}$ b) maximum $\varepsilon^E_{xx-Max.}$, c) minimum $\varepsilon^E_{xx-Min.}$, and d) standard deviation $\varepsilon^E_{xx-STD}$  of the distribution of elastic strain within the \textbf{duplicate} experimental X-ray diffraction volume with respect to time $t$ extracted using trained GPR surrogate models. The red error bars correspond to the square root of the variance (standard deviation) of the GPR surrogate model predictions.}
  \label{fig:dup_exx}
\end{figure}

\begin{figure}[h!]
  \centering \includegraphics[width=1.0\textwidth]{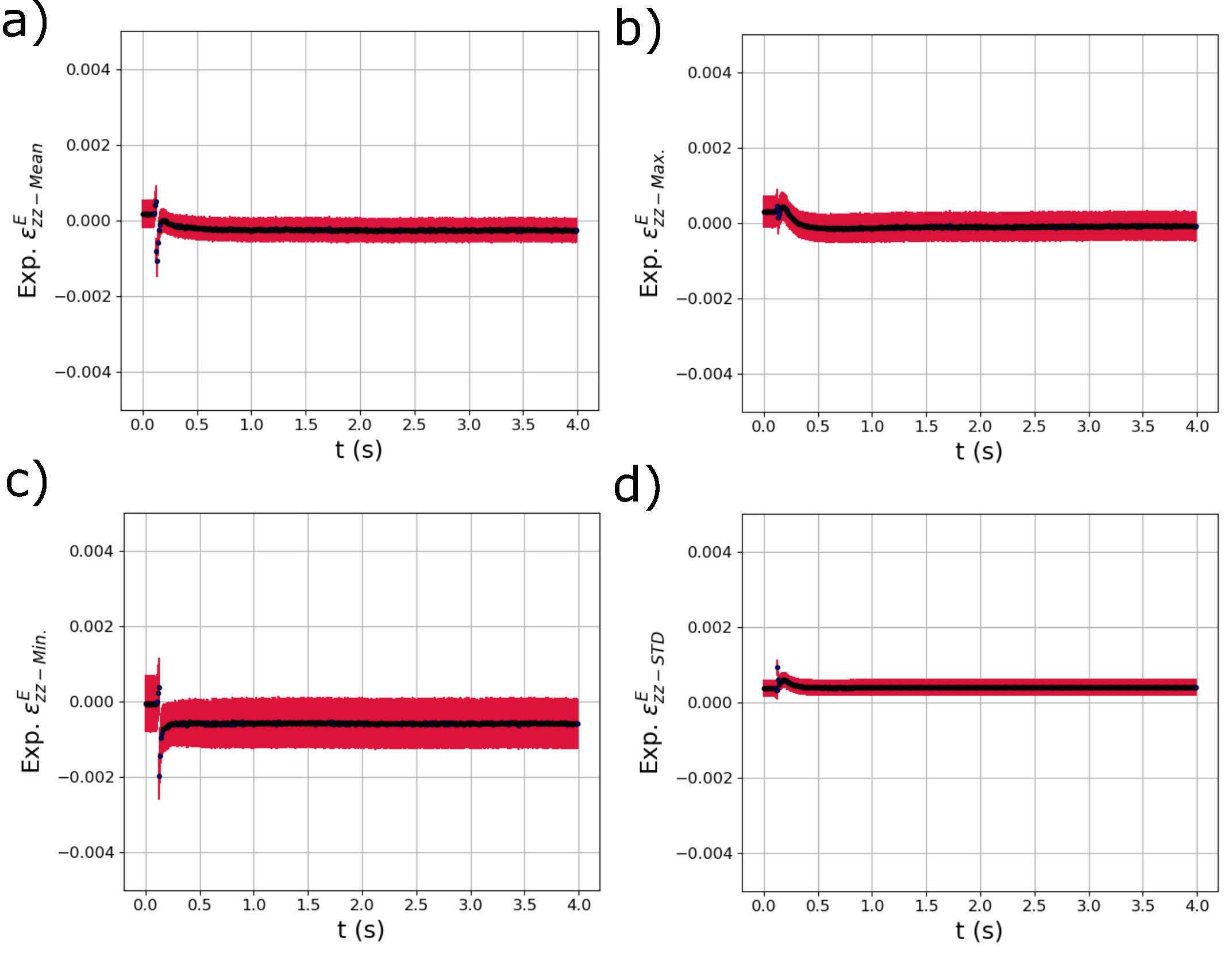}
  \caption{The evolving a) mean $\varepsilon^E_{zz-Mean}$ b) maximum $\varepsilon^E_{zz-Max.}$, c) minimum $\varepsilon^E_{zz-Min.}$, and d) standard deviation $\varepsilon^E_{zz-STD}$  of the distribution of elastic strain within the \textbf{duplicate} experimental X-ray diffraction volume with respect to time $t$ extracted using trained GPR surrogate models. The red error bars correspond to the square root of the variance (standard deviation) of the GPR surrogate model predictions.}
  \label{fig:dup_ezz}
\end{figure}

\begin{figure}[h!]
  \centering \includegraphics[width=1.0\textwidth]{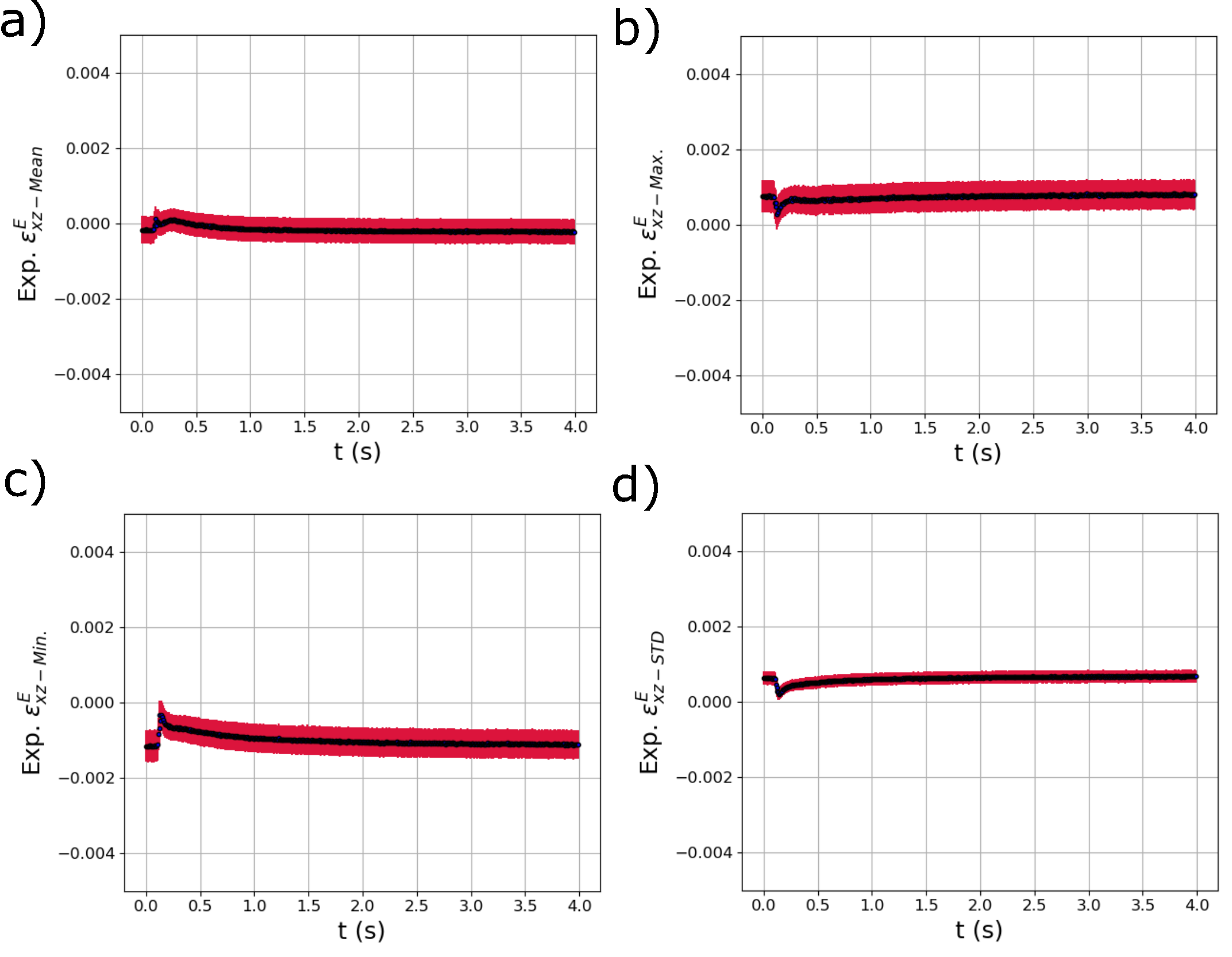}
  \caption{The evolving a) mean $\varepsilon^E_{xz-Mean}$ b) maximum $\varepsilon^E_{xz-Max.}$, c) minimum $\varepsilon^E_{xz-Min.}$, and d) standard deviation $\varepsilon^E_{xz-STD}$  of the distribution of elastic strain within the \textbf{duplicate} experimental X-ray diffraction volume with respect to time $t$ extracted using trained GPR surrogate models. The red error bars correspond to the square root of the variance (standard deviation) of the GPR surrogate model predictions.}
  \label{fig:dup_exz}
\end{figure}

After the initial laser pass over the wall specimen from which results were presented in \S \ref{sec:results}, the specimen was allowed to nominally cool and then was translated approximately 2 mm with respect to the laser and incoming X-ray beam along $\bm{x}$ such that unmelted material was now illuminated. The previously described \emph{in situ} laser melting experiment was then repeated with the same X-ray beam, detector, and laser parameters. The diffraction images from this measurement were then processed identically to that described previously and used as input for the trained GPR models. Figure \ref{fig:dup_et} shows the predicted $\varepsilon^T$ metrics, Fig. \ref{fig:dup_exx} shows $\varepsilon^E_{xx}$ metrics, Fig. \ref{fig:dup_ezz} shows $\varepsilon^E_{zz}$ metrics, and Fig. \ref{fig:dup_exz} shows $\varepsilon^E_{xz}$ metrics. As can be seen, the evolution of the $\varepsilon^T$, $\varepsilon^E_{xx}$, and $\varepsilon^E_{zz}$ are very similar to that reported in \S \ref{sec:results}, providing confidence in the repeatability of the measurements and data analysis approach. There are differences in the distribution metrics output for $\varepsilon^E_{xz}$, particularly as the laser passed over the diffraction volume, but this is likely related to the shear strains being the most sensitive to any minor misalignment of the laser path with the wall specimen geometry.

\end{document}